\documentclass[journal]{IEEEtran}
\usepackage{stfloats}
\usepackage{cite}
\usepackage{amsmath,amssymb,amsfonts}
\usepackage{graphicx}
\usepackage{textcomp}
\usepackage{xcolor}
\def\BibTeX{{\rm B\kern-.05em{\sc i\kern-.025em b}\kern-.08em
    T\kern-.1667em\lower.7ex\hbox{E}\kern-.125emX}}
\usepackage{amsmath}   
\usepackage{verbatim}
\usepackage{flushend}

\usepackage[utf8]{inputenc}

\usepackage{epstopdf}
\usepackage{stmaryrd}
\usepackage{multirow,url}
\usepackage{subfigure}
\usepackage{enumerate}
\usepackage{amsmath}
\usepackage{amsthm}

\ifodd 1
\newcommand{\com}[1]{\textbf{\color{red} (COMMENT: #1)}} 
\newcommand{\comg}[1]{\textbf{\color{green} (COMMENT: #1)}}
\newcommand{\response}[1]{\textbf{\color{magenta} (RESPONSE: #1)}} 
\else

\newcommand{\com}[1]{}
\newcommand{\comg}[1]{}
\newcommand{\response}[1]{}
\fi

\begin{document}

\title{Timely Status Update in Relay-Assisted Cooperative Communications}

\author{Haoyuan~Pan,~\IEEEmembership{Member,~IEEE,}~Jian~Feng,~\IEEEmembership{Graduate Student Member,~IEEE,}~Tse-Tin~Chan,~\IEEEmembership{Member,~IEEE,}\\~Victor~C.~M.~Leung,~\IEEEmembership{Life Fellow,~IEEE,}~Jianqiang~Li,~\IEEEmembership{Member,~IEEE}%

\thanks{H. Pan, J. Feng, Victor C. M. Leung, and J. Li are with the College of Computer Science and Software Engineering, Shenzhen University, Shenzhen, China (e-mails: {hypan@szu.edu.cn}, {fengjian2020@email.szu.edu.cn}, {vleung@ieee.org}, {lijq@szu.edu.cn}). }
\thanks{T.-T.~Chan is with the Department of Mathematics and Information Technology, The Education University of Hong Kong, Hong Kong SAR, China (e-mail: {tsetinchan@eduhk.hk}).}
}

\maketitle

\begin{abstract}
We investigate the age of information (AoI) of a relay-assisted cooperative communication system, where a source node sends status update packets to the destination node as timely as possible with the aid of a relay node. For time-slotted systems without relaying, prior works have shown that the source should generate and send a new packet to the destination every time slot to minimize the average AoI, regardless of whether the destination has successfully decoded the packet in the previous slot. However, when a dedicated relay is involved, whether the relay can improve the AoI performance requires an in-depth study. In particular, the packet generation and transmission strategy of the source should be carefully designed to cooperate with the relay. Depending on whether the source and the relay are allowed to transmit simultaneously, two relay-assisted schemes are investigated: time division multiple access (TDMA) and non-orthogonal multiple access (NOMA) schemes. In TDMA, the source generates and sends a new packet \emph{every other time slot} to avoid possible simultaneous transmission with the relay. In NOMA, the source generates and sends a new packet \emph{every time slot}, thus possibly forming simultaneous transmission from the relay and the source. A key challenge in deriving their theoretical average AoI is that the destination has different probabilities of successfully receiving an update packet in different time slots. We model each scheme using a Markov chain to derive the corresponding closed-form average AoI. Interestingly, our theoretical analysis indicates that the relay-assisted schemes can only outperform the non-relay scheme in average AoI when the signal-to-noise ratio of the source-destination link is below $-2dB$. Furthermore, comparing the merits of relay-assisted schemes, simulation results show that the TDMA scheme has a lower energy consumption, while the NOMA counterpart typically achieves a lower average AoI.
\end{abstract}



\section{Introduction}
\label{section-intro}
In recent years, timely status update systems have received significant attention from academia and industry \cite{Yates2021,Zheng2020,Gu2019}. Enabled by the Internet of Things (IoT) technology, machine-type devices with sensing, monitoring, and communication capabilities send status packets to update their latest information at the sink. Providing timely and fresh status updates is paramount in many applications, such as real-time decision making in cyber-physical systems \cite{ZWang2019} and periodic safety message exchange in vehicular ad hoc networks \cite{Leng2019}. 

Prior works showed that metrics in conventional communication systems, such as throughput and delay, are inadequate to measure the timeliness or freshness of updates \cite{Yates2021}. Hence, age of information (AoI) was proposed to characterize the information freshness \cite{Kaul2011, Kaul2012}. Unlike delay, which measures the time required to deliver a packet successfully to the destination, AoI measures the time elapsed since the generation of the most recently received status update from a destination perspective \cite{Yates2021,Kaul2012}. More precisely, if the latest update packet received by the destination is generated by the source at time $t'$, then the instantaneous AoI at time $t$ is $t-t'$. Since its introduction in \cite{Kaul2011, Kaul2012}, AoI has been studied in various communication network settings, revealing that it is fundamentally different from delay and throughput which capture the effectiveness of data collection and transmission only \cite{Lou2021,Abdel-Aziz2020,Mankar2021}. 

\begin{figure}
\centerline{\includegraphics[width=0.28\textwidth]{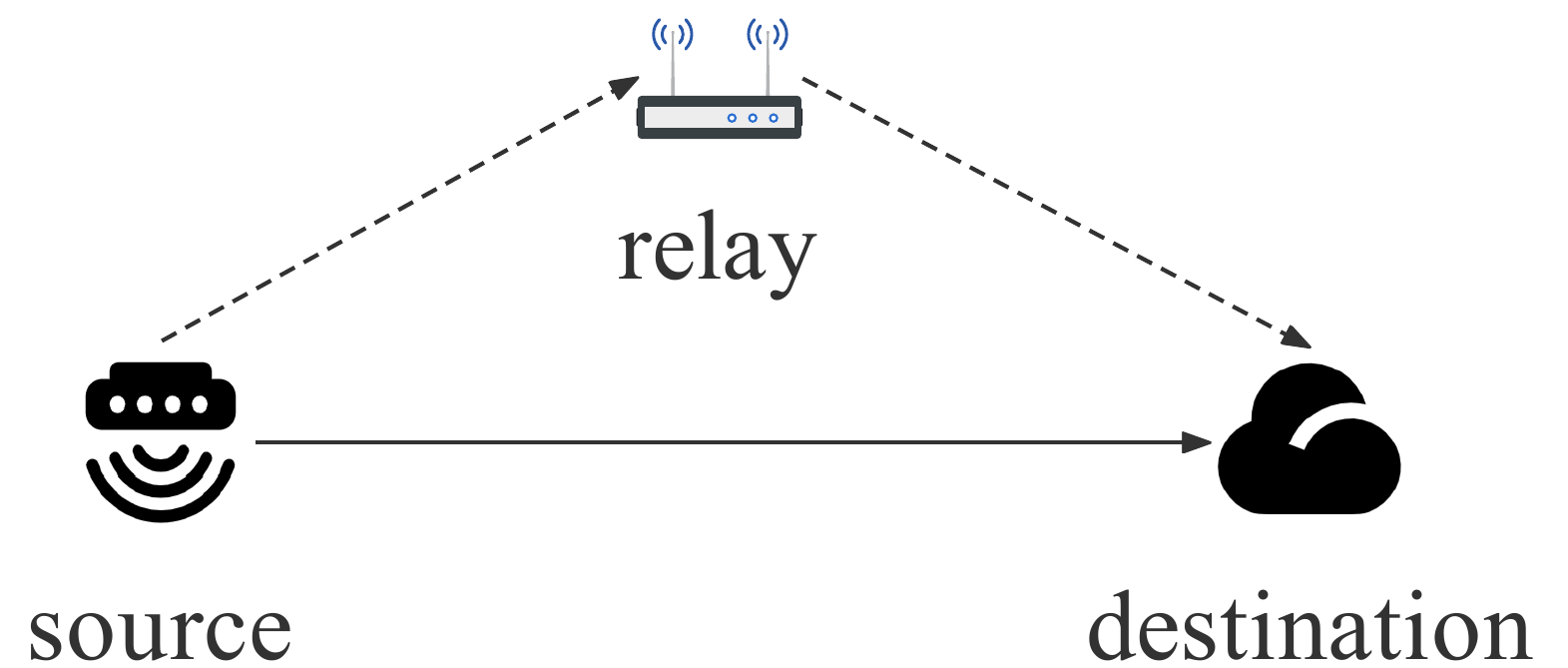}}
\caption{A relay-assisted status update system with a source node, a relay node, and a destination node. The destination may receive an update packet from the source either from the single-hop direct link (the solid line) or the two-hop relay-assisted link (the dashed lines).}
\label{fig-model}
\end{figure}

In many practical scenarios, an information source (say, a low-cost sensor) has a weak channel condition at the destination, e.g., a low signal-to-noise ratio (SNR), possibly due to a long distance from the destination or low transmit power. A dedicated relay can be used to help forward update packets of the source to the destination. In this paper, we consider a relay-assisted status update system with a source node, a half-duplex decode-and-forward (DF) relay node, and a destination node. As depicted in Fig. \ref{fig-model}, the destination may receive an update packet from the source either from the direct link or the relay-assisted link. Note that the relay-assisted link has two hops, while the direct link has one hop only. 

Intuitively, suppose a relay is located in the middle between the source and the destination (see Fig. \ref{fig-model}), the channel conditions of both hops are more likely to be better than that of the direct link, thus improving the packet reception reliability. However, it is not so obvious whether the dedicated relay can improve the information freshness. Let us consider a time-slotted system in which the transmission of an update packet occupies one time slot. If the relay is not available, i.e., a non-relay system, a ``new'' packet should always be sent by the source in each time slot regardless of whether the destination has successfully received the previous ``old'' packet. In other words, the old packet should be dropped even when it was not received by the destination, because it contains an outdated status \cite{Feng2022}. By doing so, the latest status update will be available whenever the destination successfully receives an update packet, i.e., the instantaneous AoI upon a successful reception is always one time slot. In contrast, considering a relay is involved, the destination may successfully receive the update packet more often due to better channel conditions of the relay-assisted link. However, when the destination receives an update packet via the relay-assisted link, the instantaneous AoI will only drop to two time slots (i.e., two hops are required). Then, we pose the first question (\textbf{Q1}) in this paper: 

\emph{\textbf{Q1}: Can the dedicated relay improve the AoI performance of a relay-assisted system with a direct link from the source to the destination, compared with the non-relay system?}

To answer \textbf{Q1}, we need to design the
protocol details of the relay-assisted system first. To simplify the overall system design,  we assume that there is no feedback from receivers (the destination and the relay) to transmitters (the source and the relay), which is favorable for the low-cost implementation of IoT status update applications \cite{Gu2019}. In addition, we adopt the widely used generate-at-will model \cite{Yates2021} to model the packet generation pattern at the source. Under the generate-at-will model, the status of the observed phenomena can be sampled by the source, and an update packet can be generated when the source has a transmission opportunity. For example, the source in the non-relay system generates and sends a new update packet to the destination at the beginning of each time slot to minimize the average AoI \cite{Feng2022}. Now suppose that a relay is involved. If the source still sends an update packet every time slot, the destination may receive wireless signals from the source and the relay simultaneously, i.e., when the relay successfully decodes the previous update packet and forwards it to the destination. Advanced techniques such as multiuser decoding \cite{V1998} are required to separate the simultaneously transmitted signals. To avoid simultaneous transmission, a possible solution is to lower the packet generation rate of the source, e.g., the source sends an update packet to the destination every other time slot.\footnote{We note that under frequency-division multiple access (FDMA), i.e., the source and the relay operate on different frequency bands, the source can send update packets every time slot. However, due to the reduced bandwidth of each node, the duration of each update packet increases, thus resulting in a high average AoI in FDMA \cite{Pan2020}. Therefore, we do not consider FDMA for low-AoI systems in this paper.} The discussion above poses the second interesting question (\textbf{Q2}) in this paper:

\emph{\textbf{Q2}: As far as AoI is concerned, what packet generation and transmission strategy should the source adopt when a dedicated relay is involved in the system?} 

Depending on whether the relay and the source are allowed to send simultaneously, this paper investigates two relay-assisted schemes, referred to as the time-division multiple access (TDMA) scheme and the non-orthogonal multiple access (NOMA) scheme, respectively. In the TDMA scheme, the source generates and sends a new update packet every other time slot  to avoid possible simultaneous transmission with the relay, while in the NOMA scheme, a new update packet is generated and sent every time slot (as in the non-relay scheme).

As a promising technique in the next-generation wireless communication networks, NOMA allows simultaneous transmission in the same time-frequency resource, thus increasing spectral efficiency compared with conventional orthogonal multiple access (OMA) schemes such as TDMA and OFDMA \cite{Tse2005,Ding2017}. In our setup, if the relay decodes a packet and forwards it to the destination, the source and the relay will send simultaneously in a non-orthogonal way because the source sends a new packet every time slot. To separate the two different update packets, we consider the widely used successive interference cancellation (SIC) decoder \cite{Ding2017} at the destination.
Since the relay-destination link typically has better channel conditions than the source-destination link, we assume that the SIC decoder first tries to decode the old packet forwarded by the relay, followed by the new packet from the source.  Thanks to SIC, the destination now has the opportunity to receive the most up-to-date status packet every time slot, which is not possible in the TDMA scheme. However, interference from the source reduces the effective SNR of the signals from the relay, which affects the decoding of the packet from the source due to the successive nature of the SIC. Therefore, a quantitative study is needed to investigate the relative merits of relay-assisted TDMA and NOMA schemes when AoI is the performance metric.

Unlike previous works that considered the relay-assisted link only \cite{Xie2021relay, Moradian2020,Zheng2021}, our work considers a relay-assisted network with both direct and relay-assisted links from the source to the destination. As a result, a key challenge in deriving the theoretical average AoI for both TDMA and NOMA schemes is that the destination has different probabilities of successfully receiving an update packet in different time slots (see Section \ref{section-OMA} for the details). In particular, for the NOMA scheme, in each time slot, the destination may receive an update packet from either the relay or the source. The instantaneous AoI upon a successful update at the destination could be different, i.e., one time slot (two time slots) if the packet is received via the direct link (the relay-assisted link). To tackle the above challenges, we model each scheme using a Markov chain so as to derive their corresponding closed-form average AoI. We design the states and state transitions in the Markov chains to  carefully distinguish the two different update scenarios (i.e., update packets received from the source or the relay). In addition, since practical status update packets in IoT devices are typically short (e.g., tens of bytes), we estimate the packet error rate (PER) of each link in the relay-assisted systems using the short packet theory \cite{Polyanskiy2010}. Our analysis serves as a guideline for the AoI analysis in relay-assisted systems.

Our theoretical analysis and simulation results indicate that for \textbf{Q1}, the relay-assisted schemes can only outperform the non-relay scheme in average AoI when the SNR from the direct link is below $-2dB$. Otherwise, a pure non-relay scheme can achieve a better average AoI in general. For the investigation of \textbf{Q2}, we find that if the relay-destination link has a higher enough SNR than the source-destination link, the source should generate and send an update packet every time slot, i.e., to adopt the relay-assisted NOMA scheme. With the help of SIC, the NOMA scheme can reduce the average AoI of its TDMA counterpart by as large as 25\%. However, when the source is low-power (e.g., tiny sensors) and the average AoI requirement is less stringent, the TDMA scheme is preferable to energy-efficient relay-assisted status update systems.

To conclude, we have the following major contributions:
\begin{itemize}\leftmargin=0in
\item [(1)] We study the average AoI of relay-assisted cooperative communication systems with both direct and relay-assisted links from the source to the destination. In particular, we investigate the packet generation and transmission strategy of the source, together with the help of a dedicated relay, aiming to improve the information freshness of the source at the destination. 
\item [(2)] We establish a Markov chain-based approach for deriving the theoretical average AoI of relay-assisted systems. Specifically, we derive the closed-form average AoI formulas for two relay-assisted schemes, namely the TDMA and NOMA schemes, each modeled by a Markov chain. We believe that our Markov chain-based approach can serve as a reference for the AoI analysis in general relay-assisted systems.
\item [(3)] We conduct comprehensive simulations to compare the average AoI among different schemes. Our results indicate that due to the existence of the direct link, the relay-assisted schemes can only outperform the non-relay scheme when the SNR of the source-destination link is as weak as $-2dB$. Even under weak source-destination links, the packet generation and transmission strategy of the source in relay-assisted systems should be carefully designed to cooperate with the dedicated relay and outperform the non-relay scheme. 
\end{itemize}

\section{Related Work} \label{relatedwork}
\subsection{Age of Information (AoI)}
As a performance metric to measure information freshness, AoI has attracted significant attention in recent years. Early works on AoI focused on analyzing the average AoI or peak AoI under different abstract queueing models \cite{Yates2021,Najm2017, Inoue2017}. Packet management plays an important role in lowering AoI \cite{Wang2019, Moltafet2021, Hu2021}. For example, the optimal packet generation rates under various models were investigated in \cite{Hu2021}. It was shown that the optimality conditions for AoI are usually different from those for delay and throughput \cite{Yates2021}. In addition, scheduling, sampling, and updating policies were designed to control packet transmission and minimize the average AoI \cite{Yates2021,Gong2020,Zhou2019}. These works focused mainly on the upper layers of the communication protocol stack, i.e., above the physical (PHY) and medium access control (MAC) layers. Along this line, our work investigates AoI-aware packet generation and transmission strategies as well, but focuses on the PHY and MAC layer designs in relay-assisted systems.

The study of AoI at the PHY and MAC layers is mainly focused on designing different techniques to deal with wireless impairments that lead to packet loss or errors, such as channel access policies \cite{Zhou2022}, automatic repeat request (ARQ) \cite{Xie2020, Feng2022, Ceran2019} and multiple antennas \cite{Yu2021}, thereby improving the information freshness. For example, the impacts of ARQ on the average AoI were investigated in different network settings, such as single-hop networks \cite{Xie2020} and multi-hop networks \cite{Feng2022}. In these works, feedback signals are usually required to issue from receivers to transmitters so that ARQ can be used. Our work assumes no feedback from receivers to transmitters, thus simplifying the overall system design. Without feedback, we show that in a relay-assisted setting, the packet generation and transmission strategy at the source should be jointly designed with the dedicated relay, which is of significant importance to achieve good AoI performance at the destination.

\subsection{Orthogonal and Non-Orthogonal Multiple Access}
NOMA is a key enabler for the 5G/B5G communication systems to accommodate massive devices and to improve network throughput \cite{Ding2017,Islam2017,Dai2015}. Compared with conventional TDMA techniques, NOMA allows more than one user to share the same resource block, thus increasing spectral efficiency and user fairness \cite{Ding2017,Islam2017,Dai2015}. The application of NOMA in relay-assisted systems has also been frequently studied, e.g., \cite{Ding2017,Wu2018}. In the relay-assisted NOMA scheme studied in this paper, the relay and the source may send different update packets simultaneously to the destination, forming a NOMA transmission. Most NOMA systems assume the use of successive interference cancellation (SIC) at the receiver \cite{Dai2015}. It was shown that the significant difference in received power among the users is critical for the operation of SIC \cite{Islam2017}. Our work further examines the effect of SIC on AoI in a relay-assisted network.

Most previous NOMA works focused on optimizing conventional metrics such as throughput and delay. Very recently, some works have applied NOMA to improve the average AoI performance \cite{Maatouk2019, Pan2021,Ren2022}. For example, \cite{Maatouk2019} studied the average AoI in a two-user NOMA network for the first time and showed that NOMA and TDMA outperform each other under different packet arrival rates. Later, \cite{Ren2022} specifically investigated the effect of SIC on the average AoI, and \cite{Pan2021} further validated the AoI improvement of NOMA over TDMA through experiments on software-defined radios. However, \cite{Maatouk2019,Pan2021,Ren2022} focused on single-hop networks without relays. \cite{Wu2022} studied a user-cooperative scenario with more than one hop, where an adaptive NOMA/OMA downlink system was proposed to minimize AoI. To the best of our knowledge, the use of NOMA in AoI-oriented relay-assisted systems has not been investigated. This paper is an attempt to fill in this gap.

\subsection{AoI in Relay-assisted Cooperative Communications}
A few works have explored AoI in relay-assisted TDMA cooperative communications, including static relays \cite{Feng2022,Xie2021relay, Moradian2020,Zheng2021,Li2021} and mobile relays \cite{Samir2022,Basnayaka2022}. Reference \cite{Feng2022} studied whether ARQ should be used at the relay node in a two-hop relay network by formulating the communication procedure as Markov chains. The authors of \cite{Xie2021relay} further explored the use of truncated ARQ in each hop to reduce AoI, and \cite{Zheng2021} optimized the transmit power of a full-duplex relay and the blocklength of update packets to solve the AoI minimization problem. However, these works considered the relay-assisted link only. Our work considers a relay-assisted network with both direct and relay-assisted links from the source to the destination, where different successful updating probabilities from the relay and the source have a significant impact on the evolution of AoI, and the Markov chains should be developed to distinguish these two update scenarios. Reference \cite{Li2021} considered both links as well, but did not study NOMA transmissions. Moreover, prior work paid more attention to the operation of the relay. They did not study how the source should collaborate with the relay to achieve a low average AoI. In contrast, this paper focuses more on the packet generation and transmission schemes at the source. Together with the aid of the relay, two relay-assisted schemes, namely TDMA and NOMA schemes, and their merits are investigated.

\section{Preliminaries}
\label{section-Pre}

\subsection{System Model and Age of Information (AoI) Metrics}
We study a relay-assisted status update system with a source node, a relay node, and a destination node, as shown in  Fig.~\ref{fig-model}. Let us consider a time-slotted system where the duration of an update packet occupies one time slot, assuming each time slot has a unit length. 
In addition, we assume that there is no feedback from the receiver (the destination or the relay) to the transmitter (the source or the relay), which is favorable for low-cost IoT status updates. In such a status update system, the destination wants to receive update packets from the source as fresh as possible, but the direct link from the source to the destination has a low SNR. Hence, a half-duplex DF relay is dedicated to helping forward the update packets of the source to the destination.\footnote{Our model can be extended to more complicated system setups, including scenarios with multiple source-destination pairs with a single relay, or multiple source nodes communicating with the same destination node via a single relay, etc. In all these scenarios, the system can be divided into different source-relay-destination groups. The different groups can operate in a TDMA manner. In each group, the TDMA or NOMA relay-assisted scheme can be adopted.}

\begin{figure}
\centerline{\includegraphics[width=0.29\textwidth]{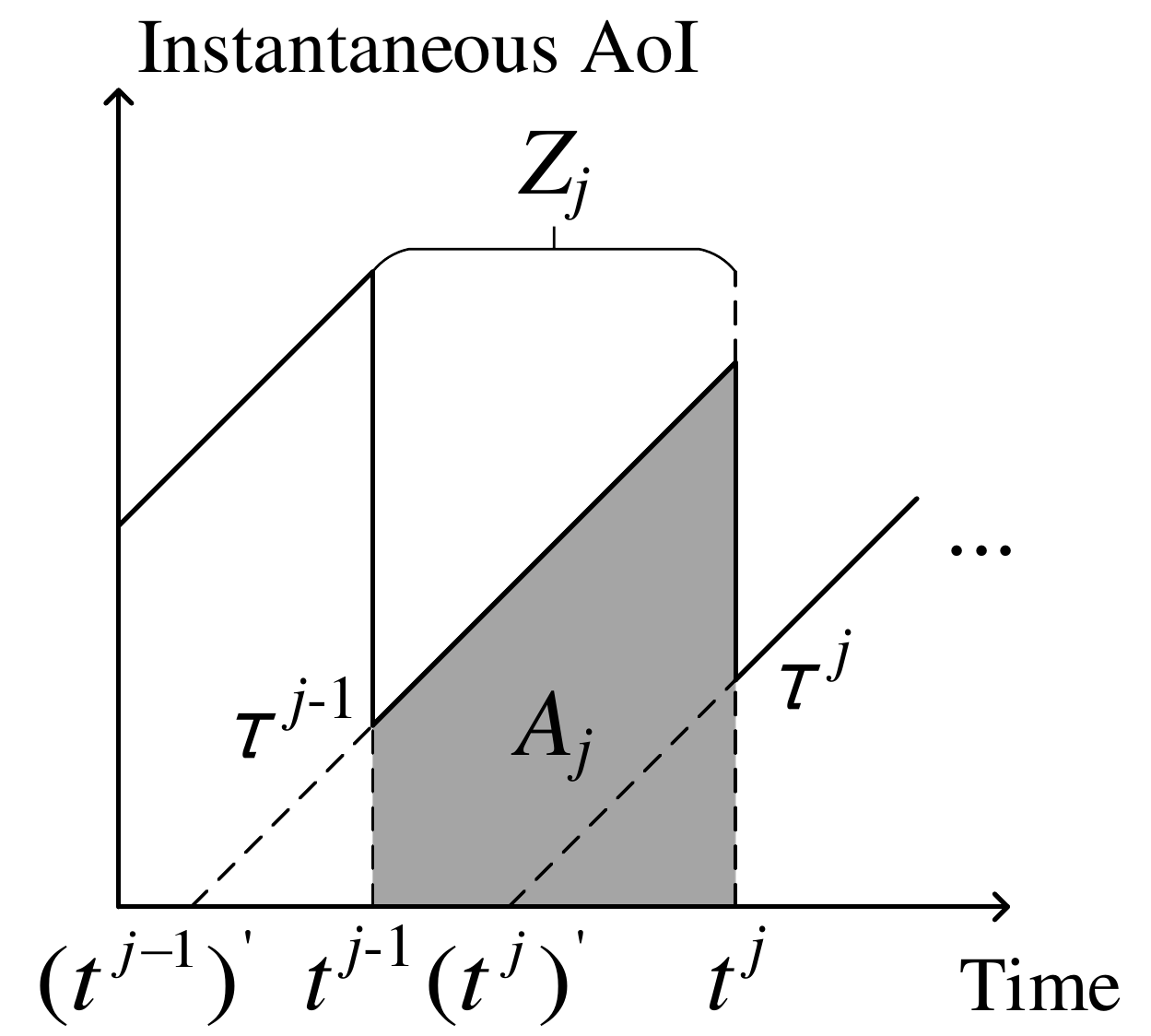}}
\caption{An example of the instantaneous AoI $\Delta (t)$. The $(j - 1)$-th and the $j$-th successful update occur at ${t^{j - 1}}$ and ${t^j}$, where the update packets are generated at ${(t^{j - 1})^{'}}$ and ${({t^{j}})^{'}}$, respectively.}
\label{fig-AoI}
\end{figure}

To measure the information freshness, we adopt the recently proposed metric, AoI, to quantify the timeliness of update packets. Specifically, at any time $t$, the instantaneous AoI of the source, measured at the destination, is defined by $\Delta (t) = t - G(t)$, \noindent where $G(t)$ is the timestamp when the most recently received update packet from the source was sampled and generated. A lower instantaneous AoI means higher information freshness. Fig.~\ref{fig-AoI} plots an example of the instantaneous AoI $\Delta (t)$, where the $(j - 1)$-th and the $j$-th successful update occur at ${t^{j - 1}}$ and ${t^j}$, respectively. Let us use $\tau $ to denote the instantaneous AoI at the moment when the destination decodes an update packet successfully. As shown in Fig.~\ref{fig-AoI}, the instantaneous AoI $\Delta (t)$ drops to ${\tau ^{j - 1}}$ and ${\tau ^j}$ at ${t^{j - 1}}$ and ${t^j}$, respectively. Between the two consecutive status updates at ${t^{j - 1}}$ and ${t^j}$, $\Delta (t)$ increases linearly with time $t$. With the instantaneous AoI, we can compute the average AoI. In particular, the average AoI of the source, $\bar \Delta $, is defined as the time average of the instantaneous AoI, i.e.,
\begin{align}
\bar \Delta  = \mathop {\lim }\limits_{T \to \infty } \frac{1}{T}\int_0^T {\Delta (t)} dt.
\label{f-average-AoI-c}
\end{align}

We consider a generate-at-will model in which the status of the observed phenomena can be sampled by the source (say, a sensor) and the update packet can be generated at any time of the sensor's choice \cite{Yates2021}.\footnote{To realize the generate-at-will model, for example, the communication layer of the source can ``pull'' a request from the upper application layer just when there is an upcoming transmission opportunity. In other words, the source should generate an update packet just before its communication layer has the transmission opportunity, not too early and not too late.} This ensures that the sampled information is as fresh as possible, e.g., a sensor reading is obtained just before the transmission opportunity. 

Under the generate-at-will model, the packet generation and transmission strategy at the source is of paramount importance to achieve low AoI in the relay-assisted system. Since there is no feedback in our considered relay-assisted system, the relay can be operated as follows. If an update packet is successfully decoded in a time slot, the relay can simply forward the update packet to the destination in the subsequent time slot. Otherwise, it remains silent or receives signals from the source (depending on whether the source sends a packet in the next time slot under the generate-at-will model). However, the packet generation and transmission scheme of the source node requires an in-depth investigation. 
Depending on whether the relay and the source are allowed to transmit simultaneously, this paper studies the relay-assisted TDMA scheme and the relay-assisted NOMA scheme, respectively, and their average AoI will be analyzed in Sections \ref{section-OMA} and \ref{section-NOMA}. Intuitively, owing to a larger packet generation rate, the NOMA scheme may outperform the TDMA scheme in average AoI at the cost of higher energy consumption. We compare the average AoI and energy cost of different schemes in Section \ref{section-Performance}. Before examining the relay-assisted scenario, we first present the general method for computing the average AoI, using the benchmarked non-relay scheme as an example.

\subsection{Review: The Average AoI of the Non-Relay Scheme}
\label{section-review}
To compute the average AoI $\bar \Delta $ in  (\ref{f-average-AoI-c}), we use $Z$ to denote the time required between two consecutive status updates. For example, in Fig.~\ref{fig-AoI}, ${Z_j}$ is the time between the $(j - 1)$-th and the $j$-th successful update. The area of ${A_j}$ between the $(j - 1)$-th and the $j$-th successful update is calculated by 
\begin{align}
A_j = {\tau ^{j - 1}}{Z_j} + \frac{1}{2}{\left( {{Z_j}} \right)^2},
\label{f-area}
\end{align}
where ${\tau ^{j - 1}}$ is the instantaneous AoI after the $(j - 1)$-th successful update. The general formula for computing the average AoI $\bar \Delta $ is given by \cite{Yates2021}
\begin{align}
\bar \Delta & = \mathop {\lim }\limits_{J \to \infty } \frac{{\sum\nolimits_{j = 1}^J {{A_j}} }}{{\sum\nolimits_{j = 1}^J {{Z_j}} }} 
= \frac{{\mathbb{E}\left[ {\tau Z + \frac{1}{2}{{\left( Z \right)}^2}} \right]}}{{\mathbb{E}\left[ Z \right]}}= \frac{\mathbb{E}\left[ \tau Z  \right]}{\mathbb{E}\left[ Z  \right]} + \frac{{\mathbb{E}\left[ {{Z^2}} \right]}}{{2\mathbb{E}\left[ Z \right]}}.
\label{f-average-AoI-s}
\end{align}
where $\mathbb{E}[\cdot]$ denotes the expectation operator. As mentioned earlier, if the relay is unavailable, the source should generate and send a new packet to the destination every time slot. If the destination successfully receives an update packet from the source, the instantaneous AoI of the source drops to $\tau =1$ (we also have $\frac{{\mathbb{E}[\tau Z]}}{{\mathbb{E}[Z]}} = 1$). Suppose that the destination can successfully decode an update packet with probability ${p_{{\rm{sd}}}}$. Then the time required for two consecutive status updates, $Z$, is a geometric random variable with the parameter ${p_{{\rm{sd}}}}$, i.e., $\mathbb{E}[Z] = 1/{p_{{\rm{sd}}}},\mathbb{E}[{Z^2}] = (2 - {p_{{\rm{sd}}}})/({p_{{\rm{sd}}}})^2$. Therefore, substituting $\mathbb{E}[Z]$ and $\mathbb{E}[{Z^2}]$ into (\ref{f-average-AoI-s}), the average AoI of the non-relay scheme is given by 
\begin{align}
{\bar \Delta _{{\rm{NR}}}} = 1 + \frac{{(2 - \;{p_{{\rm{sd}}}})/{{\left( {{p_{{\rm{sd}}}}} \right)}^2}}}{{2(1/{p_{{\rm{sd}}}})}} = \frac{1}{2} + \frac{1}{{{p_{{\rm{sd}}}}}}.
\label{f-AoI-NR}
\end{align}
From (\ref{f-AoI-NR}), we observe that if the channel condition of the direct link from the source to the destination is weak (i.e., a small ${p_{{\rm{sd}}}}$), the average AoI of the source is high because of a long time between two consecutive updates. 

In the next section, we first consider the relay-assisted TDMA scheme, where the source sends an update packet to the destination every other time slot (i.e., avoiding possible interference from the relay). A key question we want to answer is whether the relay-assisted scheme can outperform the non-relay scheme, because with the additional hop compared to the direct link transmission, the destination requires an extra time slot to receive the update packet from the source via the relay. 
We theoretically analyze the average AoI of the relay-assisted TDMA scheme. Later in Section \ref{section-NOMA}, we extend the TDMA scheme to the advanced NOMA scheme. 

\section{The Relay-assisted TDMA Scheme}
\label{section-OMA}
This section presents the relay-assisted TDMA scheme. In Section \ref{section-OMA-A}, we first describe the details of the TDMA transmission protocol and point out the difficulty in analyzing the average AoI under this scheme. Later in Section \ref{section-OMA-B}, we construct a Markov chain to model the relay-assisted TDMA scheme, from which the average AoI can be computed. As will be seen in Section \ref{section-OMA}, our analytical method also applies to the NOMA scheme with modifications to the states in the Markov chain. 

\subsection{The Relay-assisted TDMA Transmission Protocol} \label{section-OMA-A}
\begin{figure}
\centerline{\includegraphics[width=0.5\textwidth]{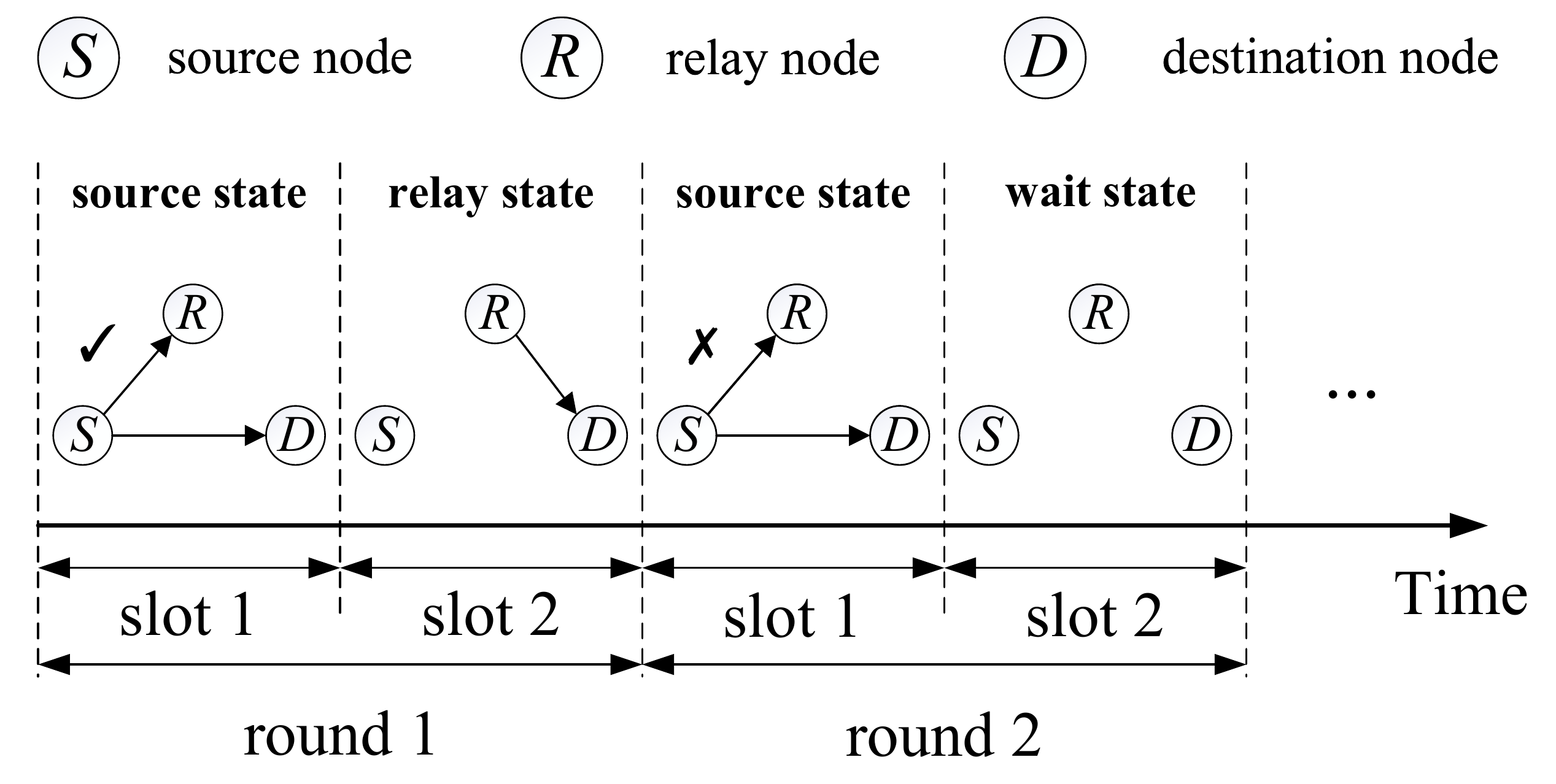}}
\caption{Example of the relay-assisted TDMA transmission procedures. }\label{fig-OMA-scheme}
\end{figure}

In the relay-assisted cooperative network operated with OMA, we assume that time is divided into multiple rounds. Each round consists of two time slots, namely slot 1 and slot 2, as shown in Fig.~\ref{fig-OMA-scheme}. We define the transmission states in each time slot as follows.
\begin{itemize}
    \item In slot 1 of a round, the source samples and sends a new update packet to the relay and the destination. We call the system in a \textbf{source state} since the source is transmitting now.
    \item In slot 2 of a round, depending on whether the relay has successfully decoded the packet from the source in slot 1, the system is either in a \textbf{relay state} or a \textbf{wait state}. More specifically, if the relay successfully receives an update packet from the source, the system now enters the \textbf{relay state} in slot 2, i.e., the relay forwards the packet to the destination in slot 2 of the current round, as shown in round 1 of Fig.~\ref{fig-OMA-scheme}. Otherwise, the system enters the \textbf{wait state}, i.e., both the source and the relay remain silent in slot 2, as shown in round 2 of Fig.~\ref{fig-OMA-scheme}.
\end{itemize}

The relay-assisted setting raises several difficulties in analyzing the average AoI. For example, the destination may receive an update packet from the source via the direct link (i.e., requiring only one time slot) or the relay-assisted link (i.e., requiring two time slots), leading to a different $\tau $ just after each successful update (recall that $\tau $ is always equal to 1 in the non-relay scheme). Furthermore, for the non-relay scheme, the decoding outcome of each time slot is independent. In each time slot, the destination decodes an update packet from the source with a success probability of ${{p_{{\rm{sd}}}}}$. However, for the relay-assisted scheme, the transmission of the relay in slot 2 depends on the decoding result in slot 1, thus leading to different success probabilities of receiving an update packet at the destination in different time slots. As a result, the time between two consecutive status updates, $Z$, is not as simple as a geometric random variable. To address this issue, as we will present in the next subsection, we model the relay-assisted TDMA scheme with the help of a Markov chain so that we can derive the average AoI.

\subsection{Average AoI of The Relay-assisted TDMA Scheme}
\label{section-OMA-B}

\begin{figure}
\centerline{\includegraphics[width=0.38\textwidth]{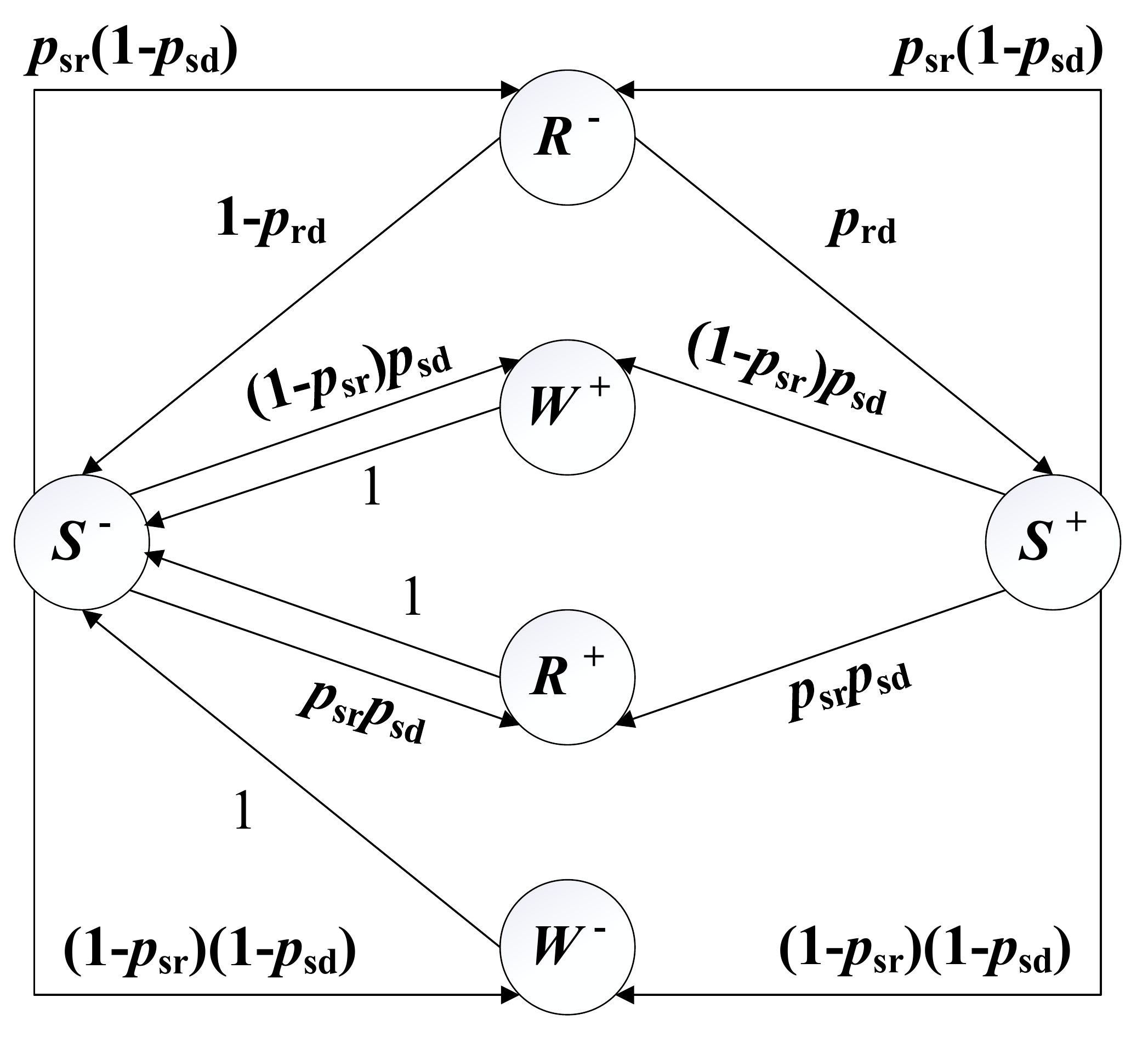}}
 \caption{The Markov chain for the relay-assisted TDMA scheme.}\label{fig-OMA-chain}
\end{figure}

We model the relay-assisted TDMA scheme using the Markov chain depicted in Fig.~\ref{fig-OMA-chain}. The Markov chain consists of six states. They are defined based on whether there is a successful update at the destination in the current time slot and the state that the system will transit in the next time slot (namely, the source state, the relay state, or the wait state; see Section \ref{section-OMA-A}):
\begin{itemize}
    \item State ${S^ + }$ (State ${S^ - }$ ) means that the destination successfully receives (fails to receive) an update packet in slot 2 of the current round. The system will enter the source state in slot 1 of a new round.
    \item State ${R^ + }$ (State ${R^ - }$) means that the destination successfully receives (fails to receive) an update packet in slot 1 of the current round. As the relay successfully receives the packet from the source in slot 1, the system will enter the relay state in slot 2.
    \item State ${W^ + }$ (State ${W^ - }$) means that the destination successfully receives (fails to receive) an update packet in slot 1 of the current round. As the relay fails to receive the packet from the source in slot 1, the system will enter the wait state in slot 2.
\end{itemize}

We use $Q$ to denote the state space, i.e., $Q = \{ {S^ + },{S^ - },{R^ + },{R^ - },{W^ + },{W^ - }\} $. Let $V = \{ {S^ + },{R^ + },{W^ + }\}$ denote the set that includes states indicating a successful update at the destination in the current time slot. In addition, suppose that the relay and the destination successfully receive an update packet from the source with probabilities of ${p_{{\rm{sr}}}}$ and ${p_{{\rm{sd}}}}$, respectively, and the destination successfully receives an update packet from the relay with probability ${p_{{\rm{rd}}}}$. With these notations, we detail the Markov chain shown in Fig.~\ref{fig-OMA-chain} as follows. 

In slot 1 of a round, the system is in either state ${S^ + }$ or state ${S^ - }$ of the Markov chain, depending on whether the previous round has a successful update at the destination in slot 2. As illustrated in Section \ref{section-OMA-A}, the source state means that the source will generate and send a new update packet to the relay and the destination. Based on the decoding results at the relay and the destination, the Markov chain may transit to state ${W^ + }$, ${W^ - }$, ${R^ + }$, or ${R^ - }$. For example, if the relay can successfully decode the packet but the destination fails to do so with probability ${p_{{\rm{sr}}}}\left( {1 - {p_{{\rm{sd}}}}} \right)$, the Markov chain transits to state ${R^ - }$. In other words, the destination fails to receive the packet in slot 1, but the relay can successfully receive it, which will be forwarded to the destination in slot 2. Likewise, if neither the relay nor the destination decodes the packet with probability $\left( {1 - {p_{{\rm{sr}}}}} \right)\left( {1 - {p_{{\rm{sd}}}}} \right)$, the Markov chain transits to state ${W^ - }$, i.e., a wait state with no packets being sent in slot 2 of the current round. 

When the current state is state ${W^ + }$, ${W^ - }$, ${R^ + }$, or ${R^ - }$, the Markov chain will go back to ${S^ + }$ or state ${S^ - }$ after slot 2, i.e., a new round starts. The state transitions are as follows.
\begin{enumerate}[1)]
    \item When the current state is state ${R^ - }$, the Markov chain will transit to state ${S^ + }$ if the destination successfully receives the packet forwarded by the relay in slot 2 with probability ${p_{{\rm{rd}}}}$.\footnote{The destination can combine the received signals in both time slots to jointly decode the update packet sent by the source, i.e., using maximum ratio combining (MRC), thus increasing ${p_{{\rm{rd}}}}$. The average AoI of the relay-assisted TDMA scheme with MRC can be found in Appendix \ref{sec:appendix_A}. As shown in Proposition 1, the average AoI is jointly affected by ${p_{{\rm{sd}}}}$, ${p_{{\rm{sr}}}}$, and ${p_{{\rm{rd}}}}$. Hence, when ${p_{{\rm{sr}}}}$ is small, increasing ${p_{{\rm{rd}}}}$ alone by MRC reduces the average AoI little when ${p_{{\rm{sr}}}}$ is small. } Otherwise, the Markov chain will transit to state ${S^ - }$ with probability $1 - {p_{{\rm{rd}}}}$.
    \item When the current state is state ${R^ + }$, i.e., both the destination and the relay receive an update packet from the source in slot 1, the system will transit to state ${S^ - }$ with probability one. This happens even though the destination may receive the packet forwarded by the relay in slot 2, because the same packet has been received in slot 1 and the destination will not decode it in slot 2. 
    \item When the current state is state ${W^ + }$ or state ${W^ - }$, the system will transit to state ${S^ - }$ with probability one, i.e., no packets are sent in the wait state and thus the destination does not receive an update packet in slot 2.
\end{enumerate}

\newcounter{mytempeqncnt}
\begin{figure*}[!t]
\setcounter{mytempeqncnt}{\value{equation}}
\footnotesize
\begin{align}
{\Omega _{{\rm{TDMA}}}} &=\begin{pmatrix}
    {{\omega _{{W^ + }{W^ + }}}} & {{\omega _{{W^ + }{R^ + }}}} & {{\omega _{{W^ + }{S^ + }}}} & {{\omega _{{W^ + }{R^ - }}}} & {{\omega _{{W^ + }{W^ - }}}} & {{\omega _{{W^ + }{S^ - }}}}\cr 
    {{\omega _{{R^ + }{W^ + }}}} & {{\omega _{{R^ + }{R^ + }}}}  & {{\omega _{{R^ + }{S^ + }}}}  & {{\omega _{{R^ + }{R^ - }}}} & {{\omega _{{R^ + }{W^ - }}}} & {{\omega _{{R^ + }{S^ - }}}}\cr 
    {{\omega _{{S^ + }{W^ + }}}} & {{\omega _{{S^ + }{R^ + }}}} & {{\omega _{{S^ + }{S^ + }}}} & {{\omega _{{S^ + }{R^ - }}}} & {{\omega _{{S^ + }{W^ - }}}} & {{\omega _{{S^ + }{S^ - }}}}\cr 
    {{\omega _{{R^ - }{W^ + }}}} & {{\omega _{{R^ - }{R^ + }}}} & {{\omega _{{R^ - }{S^ + }}}} & {{\omega _{{R^ - }{R^ - }}}} & {{\omega _{{R^ - }{W^ - }}}} & {{\omega _{{R^ - }{S^ - }}}}\cr 
    {{\omega _{{W^ - }{W^ + }}}} & {{\omega _{{W^ - }{R^ + }}}}  & {{\omega _{{W^ - }{S^ + }}}}  & {{\omega _{{W^ - }{R^ - }}}} & {{\omega _{{W^ - }{W^ - }}}} & {{\omega _{{W^ - }{S^ - }}}}\cr 
    {{\omega _{{S^ - }{W^ + }}}} & {{\omega _{{S^ - }{R^ + }}}} & {{\omega _{{S^ - }{S^ + }}}} & {{\omega _{{S^ - }{R^ - }}}} & {{\omega _{{S^ - }{W^ - }}}} & {{\omega _{{S^ - }{S^ - }}}}\cr 
\end{pmatrix} \cr
&=\begin{pmatrix}
    0 & 0 & 0 & 0 & 0 & 1 \cr
    0 & 0 & 0 & 0 & 0 & 1 \cr
    {\left( {1 - {p_{{\rm{sr}}}}} \right){p_{{\rm{sd}}}}} & {{p_{{\rm{sr}}}}{p_{{\rm{sd}}}}} & 0 & {{p_{{\rm{sr}}}}\left( {1 - {p_{{\rm{sd}}}}} \right)} & {\left( {1 - {p_{{\rm{sr}}}}} \right)\left( {1 - {p_{{\rm{sd}}}}} \right)} & 0 \cr 
    0 & 0 & {{p_{{\rm{rd}}}}} & 0 & 0 & {1 - {p_{{\rm{rd}}}}} \cr 
    0 & 0 & 0 & 0 & 0 & 1 \cr
    {\left( {1 - {p_{{\rm{sr}}}}} \right){p_{{\rm{sd}}}}} & {{p_{{\rm{sr}}}}{p_{{\rm{sd}}}}} & 0 & {{p_{{\rm{sr}}}}\left( {1 - {p_{{\rm{sd}}}}} \right)} & {\left( {1 - {p_{{\rm{sr}}}}} \right)\left( {1 - {p_{{\rm{sd}}}}} \right)} & 0 \cr  
\end{pmatrix}.
\label{f-matrix-OMA}
\end{align}
\hrulefill
\end{figure*}

Let ${\Omega _{{\rm{TDMA}}}}$ denote the state transition matrix of the Markov chain shown in Fig.~\ref{fig-OMA-chain}. Then ${\Omega _{{\rm{TDMA}}}}$ can be written as (\ref{f-matrix-OMA}),
where ${\omega _{xy}}$ is the transition probability from state $J=x$ to state $J=y$ for any $x,y \in Q$. Using the Markov chain, we have the following Proposition 1.

\textbf{Proposition 1:} The average AoI of the relay-assisted TDMA transmission scheme, ${\bar \Delta _{{\rm{TDMA}}}}$, is given by
\begin{align}
{\bar \Delta _{{\rm{TDMA}}}} = \frac{{\mathbb{E}\left[ {\tau Z} \right]}}{{\mathbb{E}\left[ Z \right]}} + \frac{{\mathbb{E}\left[ {{Z^2}} \right]}}{{2\mathbb{E}\left[ Z \right]}} = 1 + \frac{{2 - {p_{{\rm{sd}}}}}}{{{p_{{\rm{sd}}}} + {p_{{\rm{sr}}}}{p_{{\rm{rd}}}}\left( {1 - {p_{{\rm{sd}}}}} \right)}}.
\label{f-aoi-OMA}
\end{align}

\begin{proof}
To compute the average AoI of the relay-assisted TDMA scheme by (\ref{f-average-AoI-s}), we need to compute $\mathbb{E}\left[ Z \right]$, $\mathbb{E}\left[ {{Z^2}} \right]$, and $\mathbb{E}\left[ {\tau Z} \right]$ accordingly. The time between two consecutive status updates, $Z$, is equal to the time required to start from state ${J_0}$ to state ${J_Z}$, where ${J_0}, {J_Z} \in V$, going through a series of intermediate states ${J_1},{J_2}, \cdots ,{J_{Z - 1}} \notin V$. Hence, $\mathbb{E}\left[ Z \right]$ can be computed by 
\begin{align}
&\mathbb{E}\left[ Z \right] = \frac{{{\pi _{{W^ + }}}}}{{{\pi _{{W^ + }}} + {\pi _{{R^ + }}} + {\pi _{{S^ + }}}}}{M_{{W^ + }V}} \notag \\
&+ \frac{{{\pi _{{R^ + }}}}}{{{\pi _{{W^ + }}} + {\pi _{{R^ + }}} + {\pi _{{S^ + }}}}}{M_{{R^ + }V}}+ \frac{{{\pi _{S + }}}}{{{\pi _{{W^ + }}} + {\pi _{{R^ + }}} + {\pi _{{S^ + }}}}}{M_{{S^ + }V}},
\label{f-E_Z-OMA-equation}
\end{align}
where ${M_{aV}}$ denotes the expected time required to traverse from state ${J_0} = a$ to state ${J_Z} \in V$ for the first time and $\pi  = \left( {{\pi _{{W^ + }}},{\pi _{{R^ + }}},{\pi _{{S^ + }}},{\pi _{{R^ - }}},{\pi _{{W^ - }}},{\pi _{{S^ - }}}} \right)\;$ is the stationary distribution of the Markov chain depicted in Fig.~\ref{fig-OMA-chain}. Similarly, $\mathbb{E}\left[ {{Z^2}} \right]$ can be computed by 
\begin{align}
&\mathbb{E}\left[ {{Z^2}} \right] = \frac{{{\pi _{{W^ + }}}}}{{{\pi _{{W^ + }}} + {\pi _{{R^ + }}} + {\pi _{{S^ + }}}}}{N_{{W^ + }V}} \notag \\
&+ \frac{{{\pi _{{R^ + }}}}}{{{\pi _{{W^ + }}} + {\pi _{{R^ + }}} + {\pi _{{S^ + }}}}}{N_{{R^ + }V}} + \frac{{{\pi _{S + }}}}{{{\pi _{{W^ + }}} + {\pi _{{R^ + }}} + {\pi _{{S^ + }}}}}{N_{{S^ + }V}},
\label{f-E_Z^2-OMA-equation}
\end{align}
where ${N_{aV}}$ is the expectation of the second moment of the time required to traverse from state ${J_0} = a$ to state ${J_Z} \in V$ for the first time.

\begin{figure*}[!t]
\setcounter{mytempeqncnt}{\value{equation}}
\footnotesize
\begin{align}
\mathbb{E}\left[ {\tau Z} \right] = 1 \cdot \left( {\frac{{{\pi _{{W^ + }}}}}{{{\pi _{{W^ + }}} + {\pi _{{R^ + }}} + {\pi _{{S^ + }}}}}{M_{{W^ + }V}} + \frac{{{\pi _{{R^ + }}}}}{{{\pi _{{W^ + }}} + {\pi _{{R^ + }}} + {\pi _{{S^ + }}}}}{M_{{R^ + }V}}} \right) + 2 \cdot \left( {\frac{{{\pi _{{S^ + }}}}}{{{\pi _{{W^ + }}} + {\pi _{{R^ + }}} + {\pi _{{S^ + }}}}}{M_{{S^ + }V}}} \right).
\label{f-E_tZ-OMA-equation}
\end{align}
\hrulefill
\end{figure*}

Now, we consider the computation of $\mathbb{E}\left[ {\tau Z} \right]$. Notice that when the Markov chain reaches state ${W^+}$ or state ${R^+}$ (i.e., the destination successfully receives an update packet in slot 1 of a round), the instantaneous AoI drops to $\tau  = 1$; when the Markov chain reaches state ${S^+}$ (i.e., the destination successfully receives an update packet in slot 2 of a round), the instantaneous AoI drops to $\tau  = 2$. Therefore, $\mathbb{E}\left[ {\tau Z} \right]$ is given by (\ref{f-E_tZ-OMA-equation}). In the following, we compute the components in (\ref{f-E_Z-OMA-equation}), (\ref{f-E_Z^2-OMA-equation}), and (\ref{f-E_tZ-OMA-equation}).

\noindent \textbf{Computation of ${\pi _{{W^ + }}},{\pi _{{R^ + }}},{\pi _{{S^ + }}}$:} According to the property of a Markov chain, the stationary distribution of the Markov chain can be computed by solving equation $\pi {\Omega _{_{{\rm{TDMA}}}}} = \pi$. We have
\begin{align}
    \pi  &= \left( {\pi _{{W^ + }}}, {\pi _{{R^ + }}}, {\pi _{{S^ + }}}, {\pi _{{R^ - }}}, {\pi _{{W^ - }}}, {\pi _{{S^ - }}} \right) \notag \\ 
  &=(\frac{{\left( {1 - {p_{{\rm{sr}}}}} \right){p_{{\rm{sd}}}}}}{2}, \frac{{{p_{{\rm{sr}}}}{p_{{\rm{sd}}}}}}{2}, \frac{{{p_{{\rm{sr}}}}{p_{{\rm{rd}}}}\left( {1 - {p_{{\rm{sd}}}}} \right)}}{2}, \frac{{{p_{{\rm{sr}}}}\left( {1 - {p_{{\rm{sd}}}}} \right)}}{2},  \notag \\ 
   &~~~~~\frac{{\left( {1 - {p_{{\rm{sr}}}}} \right)\left( {1 - {p_{{\rm{sd}}}}} \right)}}{2}, \frac{{1 - {p_{{\rm{sr}}}}{p_{{\rm{rd}}}}\left( {1 - {p_{{\rm{sd}}}}} \right)}}{2} ). 
    \label{f-pi-OMA}
\end{align}

\noindent \textbf{Computation of ${M_{aV}}$ and ${N_{aV}}$:} Since ${M_{aV}}$ is the expected time required to traverse from state ${J_0} = a$ to ${J_Z} \in V$ for the first time, it is computed by 
\begin{align}
{M_{aV}} &= \mathbb{E}\left[ {{T_V}\left| {{J_0} = a} \right.} \right] \notag \\ & = \sum\limits_{b \in V} {\mathbb{E}\left[ {1\left| {{J_1} = b} \right.} \right]\Pr \left( {{J_1} = b\left| {{J_0} = a} \right.} \right)}  \notag \\
&+ \sum\limits_{b \notin V} {\mathbb{E}\left[ {1 + {T_V}\left| {{J_1} = b} \right.} \right]\Pr \left( {{J_1} = b\left| {{J_0} = a} \right.} \right)} {\rm{ }},
\label{f-M}
\end{align}
where ${T_V}$ is a random variable that represents the time to reach state space $V$ for the ﬁrst time. The first term of (\ref{f-M}) means that starting from ${J_0} = a$ and after one step of state transition, the Markov chain reaches $V$ for the first time. The second term of (\ref{f-M}) means that starting from ${J_0} = a$ and after one step of state transition, the next state does not belong to $V$, and it requires additional ${T_V}$ state transitions to reach $V$ for the first time. Based on the Markov chain in Fig.~\ref{fig-OMA-chain}, we can write the expressions for the different ${M_{aV}}$ accordingly, forming a set of linear equations 
\begin{align}
\left\{ \begin{array}{l}
{M_{{W^ + }V}} = 1 + {M_{{S^ - }V}}\\
{M_{{R^ + }V}} = 1 + {M_{{S^ - }V}}\\
{M_{{S^ + }V}} = {\omega _{{S^ + }{W^ + }}} + {\omega _{{S^ + }{R^ + }}} + {\omega _{{S^ + }{R^ - }}}(1 + {M_{{R^ - }V}}) \\
 ~~~~~~~~~~~ +{\omega _{{S^ + }{W^ - }}}(1 + {M_{{W^ - }V}})\\
{M_{{R^ - }V}} = {\omega _{{R^ - }{S^ + }}} + {\omega _{{R^ - }{S^ - }}}(1 + {M_{{S^ - }V}})\\
{M_{{W^ - }V}} = 1 + {M_{{S^ - }V}}\\
{M_{{S^ - }V}} = {\omega _{{S^ - }{W^ + }}} + {\omega _{{S^ - }{R^ + }}} + {\omega _{{S^ - }{R^ - }}}(1 + {M_{{R^ - }V}}) \\
~~~~~~~~~~~ +{\omega _{{S^ - }{W^ - }}}(1 + {M_{{W^ - }V}})
\end{array} \right..
\label{f-M-OMA-equation}
\end{align}
Solving these linear equations, we obtain ${M_{aV}}, \ a = \{ {S^ + },{R^ + },{W^ + }\} $, where
\begin{align}
\left\{ \begin{array}{l}
{M_{{S^ + }V}} = \frac{{2 - {p_{{\rm{sd}}}}}}{{{p_{{\rm{sd}}}} + {p_{{\rm{sr}}}}{p_{{\rm{rd}}}}\left( {1 - {p_{{\rm{sd}}}}} \right)}}\\
{M_{{W^ + }V}} = {M_{{R^ + }V}} = 1 + {M_{{S^ + }V}}
\end{array} \right..
\label{f-M-OMA}
\end{align}

Similar to ${M_{aV}}$, ${N_{aV}}$ can be computed by 
\begin{align}
&{N_{aV}} = \mathbb{E}\left[ {{{\left( {{T_V}} \right)}^2}\left| {{J_0} = a} \right.} \right] \notag \\
& = \sum\limits_{b \in V} {\mathbb{E}\left[ {1\left| {{J_1} = b} \right.} \right]\Pr \left( {{J_1} = b\left| {{J_0} = a} \right.} \right)}  \notag \\
&~~~ + \sum\limits_{b \notin V} {\mathbb{E}\left[ {{{(1 + {T_V})}^2}\left| {{J_1} = b} \right.} \right]\Pr \left( {{J_1} = b\left| {{J_0} = a} \right.} \right)} \notag \\
&= \sum\limits_{b \in V} {\mathbb{E}\left[ {1\left| {{J_1} = b} \right.} \right]\Pr \left( {{J_1} = b\left| {{J_0} = a} \right.} \right)}  \notag \\
&~~~ + \sum\limits_{b \notin V} {\mathbb{E}\left[ {{{({T_V})}^2} + 2{T_V} + 1\left| {{J_1} = b} \right.} \right]\Pr \left( {{J_1} = b\left| {{J_0} = a} \right.} \right)} {\rm{  }}.
\label{f-N}
\end{align}
${N_{aV}}$ can be expressed by a set of linear equations for different $a$, i.e.,
\begin{align}
\small
\left\{ \begin{array}{l}
{N_{{W^ + }V}} = 1 + {N_{{S^ - }V}} + 2{M_{{S^ - }V}}\\
{N_{{R^ + }V}} = 1 + {N_{{S^ - }V}} + 2{M_{{S^ - }V}}\\
{N_{{S^ + }V}} = {\omega _{{S^ + }{W^ + }}} + {\omega _{{S^ + }{R^ + }}} + {\omega _{{S^ + }{R^ - }}}(1 + {N_{{R^ - }V}} + 2{M_{{R^ - }V}}) \\
~~~~~~~~~~~ + {\omega _{{S^ + }{W^ - }}}(1 + {N_{{W^ - }V}} + 2{M_{{W^ - }V}})\\
{N_{{R^ - }V}} = {\omega _{{R^ - }{S^ + }}} + {\omega _{{R^ - }{S^ - }}}(1 + {N_{{S^ - }V}} + 2{M_{{S^ - }V}})\\
{N_{{W^ - }V}} = 1 + {N_{{S^ - }V}} + 2{M_{{S^ - }V}}\\
{N_{{S^ - }V}} = {\omega _{{S^ - }{W^ + }}} + {\omega _{{S^ - }{R^ + }}} + {\omega _{{S^ - }{R^ - }}}(1 + {N_{{R^ - }V}} + 2{M_{{R^ - }V}}) \\ 
~~~~~~~~~~~  + {\omega _{{S^ - }{W^ - }}}(1 + {N_{{W^ - }V}} + 2{M_{{W^ - }V}})
\end{array} \right..
\label{f-N-OMA-equation}
\end{align}
Again, solving the linear equations, we obtain ${N_{aV}}, \ a = \{ {S^ + },{R^ + },{W^ + }\} $ 
\begin{align}
\left\{ \begin{array}{l}
{N_{{S^ + }V}} = \frac{{4\left( {1 - {p_{{\rm{sd}}}}} \right)\left( {1 - {p_{{\rm{sr}}}}{p_{{\rm{rd}}}}} \right)\left( {2 - {p_{{\rm{sd}}}}} \right)}}{{{{\left[ {{p_{{\rm{sd}}}} + {p_{{\rm{sr}}}}{p_{{\rm{rd}}}}\left( {1 - {p_{{\rm{sd}}}}} \right)} \right]}^2}}} + \frac{{4 - 3{p_{{\rm{sd}}}}}}{{{p_{{\rm{sd}}}} + {p_{{\rm{sr}}}}{p_{{\rm{rd}}}}\left( {1 - {p_{{\rm{sd}}}}} \right)}}\\
{N_{{W^ + }V}} = {N_{{R^ + }V}} = 1 + {N_{{S^ + }V}} + 2{M_{{S^ + }V}}
\end{array} \right. .
\label{f-N_OMA}
\end{align}
Then $\mathbb{E}\left[ Z \right]$ is obtained by substituting (\ref{f-pi-OMA}) and (\ref{f-M-OMA}) into (\ref{f-E_Z-OMA-equation}), i.e.,
\begin{align}
\mathbb{E}\left[ Z \right] = \frac{2}{{{p_{{\rm{sd}}}} + {p_{{\rm{sr}}}}{p_{{\rm{rd}}}}\left( {1 - {p_{{\rm{sd}}}}} \right)}}.
\label{f-E_Z-OMA}
\end{align}

Similarly, we obtain $\mathbb{E}\left[ {\tau Z} \right]$ and $\mathbb{E}\left[ {{Z^2}} \right]$
\begin{align}
\mathbb{E}\left[ {\tau Z} \right] &= \frac{2}{{{p_{{\rm{sd}}}} + {p_{{\rm{sr}}}}{p_{{\rm{rd}}}}\left( {1 - {p_{{\rm{sd}}}}} \right)}} +\frac{{{p_{{\rm{sr}}}}{p_{{\rm{rd}}}}\left( {1 - {p_{{\rm{sd}}}}} \right)\left( {2 - {p_{{\rm{sd}}}}} \right)}}{{{{\left[ {{p_{{\rm{sd}}}} + {p_{{\rm{sr}}}}{p_{{\rm{rd}}}}\left( {1 - {p_{{\rm{sd}}}}} \right)} \right]}^2}}}.
\label{f-E_tZ-OMA}\\
\mathbb{E}\left[ {{Z^2}} \right] &= \frac{{2\left( {2 - {p_{{\rm{sd}}}}} \right)}}{{{p_{{\rm{sd}}}} + {p_{{\rm{sr}}}}{p_{{\rm{rd}}}}\left( {1 - {p_{{\rm{sd}}}}} \right)}} \notag \\
&+\frac{{2\left( {2 - {p_{{\rm{sd}}}}} \right)\left[ {{p_{{\rm{sd}}}} + 2\left( {1 - {p_{{\rm{sr}}}}{p_{{\rm{rd}}}}} \right)\left( {1 - {p_{{\rm{sd}}}}} \right)} \right]}}{{{{\left[ {{p_{{\rm{sd}}}} + {p_{{\rm{sr}}}}{p_{{\rm{rd}}}}\left( {1 - {p_{{\rm{sd}}}}} \right)} \right]}^2}}}.
\label{f-E_Z^2-OMA}
\end{align}
Finally, we compute ${\bar \Delta _{{\rm{TDMA}}}}$ by substituting (\ref{f-E_Z-OMA}), (\ref{f-E_tZ-OMA}) and (\ref{f-E_Z^2-OMA}) into (\ref{f-aoi-OMA}). 
\end{proof}

\section{The Relay-assisted NOMA Scheme}
\label{section-NOMA} 
This section presents the relay-assisted NOMA scheme. As in Section \ref{section-OMA}, we first describe the details of the NOMA transmission protocol and then compute its average AoI by a Markov chain, which is constructed similarly to the one in the TDMA scheme.

\subsection{The Relay-assisted NOMA Transmission Protocol}
\label{section-NOMA-A}

\begin{figure}
\centerline{\includegraphics[width=0.5\textwidth]{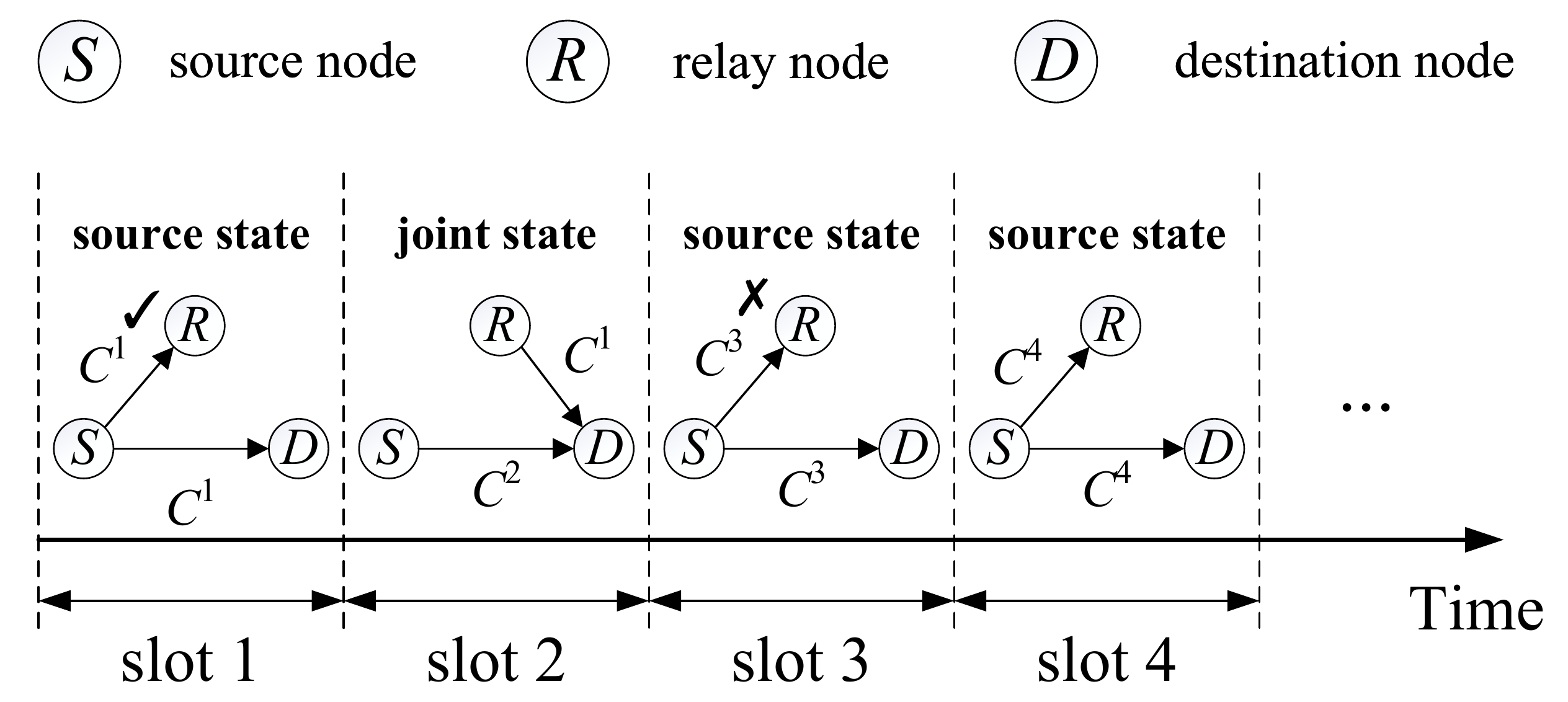}}
  \caption{Examples of the relay-assisted NOMA transmission procedures.}\label{fig-NOMA-scheme}
\end{figure}

Recall that a key difference between the NOMA scheme and the TDMA scheme is that the source sends a new update packet to the destination every time slot. Suppose that the source generates and sends a new update packet ${C^i}$ in time slot $i$, as shown in Fig.~\ref{fig-NOMA-scheme}. In the first time slot of Fig.~\ref{fig-NOMA-scheme}, the relay successfully decodes packet ${C^1}$ sent by the source. Then in the second time slot, the relay forwards the \textit{old} packet ${C^1}$ to the destination. At the same time, the source sends a \textit{new} packet ${C^2}$ to the destination, i.e., a NOMA transmission. If the relay cannot decode the packet sent by the source (e.g., the third time slot in Fig.~\ref{fig-NOMA-scheme}), it keeps silent in the subsequent time slot, and only the source sends a new packet (e.g., the fourth time slot in Fig.~\ref{fig-NOMA-scheme}).

Hence, based on whether the relay and the source send simultaneously, we define the following two transmission states in each time slot
\begin{itemize}
    \item If only the source sends a new update packet, we call the system in the \textbf{source state}.
    \item If the relay and the source send simultaneously, we call the system in the \textbf{joint state}. More specifically, the relay forwards the old packet received in the previous time slot, and the source sends a new packet, thus forming a NOMA transmission. 
\end{itemize}

As in the TDMA scheme, we also construct a Markov chain to compute the average AoI of the relay-assisted NOMA scheme in the following subsection.

\subsection{Average AoI of The Relay-assisted NOMA Scheme}
\label{section-NOMA-B}
\begin{figure}
\centerline{\includegraphics[width=0.45\textwidth]{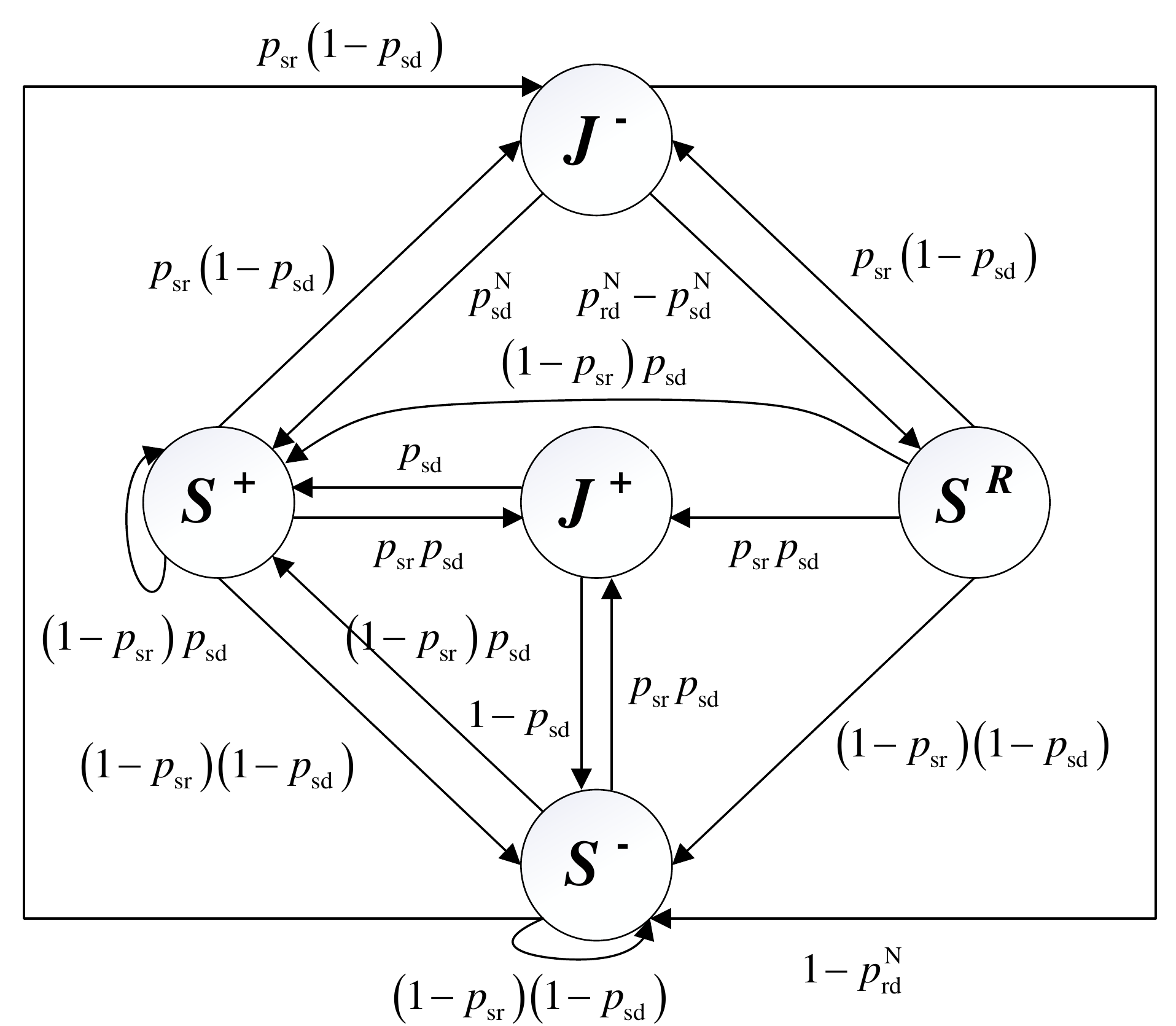}}
\caption{The Markov chain for the relay-assisted NOMA scheme.}
\label{fig-NOMA-chain}
\end{figure}

We model the relay-assisted NOMA scheme using the Markov chain depicted in Fig.~\ref{fig-NOMA-chain}. The Markov chain consists of five states. As in the TDMA scheme, the states are defined based on whether there is a successful update at the destination in the current time slot and the state of the system in the next time slot (i.e., the source state or the joint state). 
\begin{itemize}
    \item State ${S^ + }$ (${S^ - }$) means that the destination successfully receives (fails to receive) an update packet \underline{from the source} in the current time slot, and the system will enter the \textbf{source state} in the next time slot.
    \item State ${J^ + }$ (${J^ - }$) means that the destination successfully receives (fails to receive) an update packet \underline{from the source} in the current time slot, and the system will enter the \textbf{joint state} in the next time slot. 
    \item State ${S^ R }$ means that the destination successfully receives an update packet \underline{from the relay} in the current time slot of NOMA transmission (but fails to decode the update packet from the source), and the system will enter the \textbf{source state} in the next time slot.
\end{itemize}

For consistency, let us also use $Q= \{ {S^ + },{J^ + },{S^R},{J^ - },{S^ - }\}$ to denote the state space, and use  $V = \{ {S^ + },{J^ + },{S^R}\}$ to include the states indicating a successful update in the current time slot. Notice that in ${S^ R }$, the instantaneous AoI drops to two time slots upon the successful update, while in ${S^ + }$ or ${J^ + }$, the instantaneous AoI drops to one time slot only.

For the NOMA transmission, we assume the use of SIC decoding at the destination. Since the relay-destination link generally has a better channel condition than the source-destination link, the SIC decoder first tries to decode the old packet forwarded from the relay by treating the new packet sent from the source as noise. If the old packet is successfully decoded, the SIC decoder subtracts it from the superimposed signals and tries to decode the new packet. Suppose that the SIC decoder decodes the old packet with a successful decoding probability of $p_{{\rm{rd}}}^{\rm{N}}$. Then the successful decoding probability of the new packet is $p_{{\rm{sd}}}^{\rm{N}} = p_{{\rm{rd}}}^{\rm{N}}{p_{{\rm{sd}}}}$, i.e., perfect SIC is assumed so that the destination tries to decode an interference-free packet from the source.

With the notations above, we now detail the Markov chain shown in Fig.~\ref{fig-NOMA-chain}. When the current state is state ${S^ + }$, ${S^ - }$, or ${S^ R }$, the system will enter the source state, meaning that the source generates and sends a new update packet to the relay and the destination. Based on the decoding results at the relay and the destination, the Markov chain may transit to state ${S^ + }$, ${S^ - }$, ${J^ + }$, or ${J^ - }$. For example, if the relay successfully decodes the packet but the destination fails to do so, the Markov chain will transit to state ${J^ - }$ with probability ${p_{{\rm{sr}}}}\left( {1 - {p_{{\rm{sd}}}}} \right)$, meaning that the destination fails to receive the packet from the source in the current time slot, and the next time slot is a NOMA transmission (i.e., the relay will forward the decoded packet). Likewise, if neither the relay nor the destination decodes the packet with probability $\left( {1 - {p_{{\rm{sr}}}}} \right)\left( {1 - {p_{{\rm{sd}}}}} \right)$, the Markov chain will transit to state ${S^ - }$, i.e., only the source sends a new update packet.

When the current state is state ${J^ + }$ or ${J^ - }$, the system will transit to the source state after the joint state, regardless of the SIC decoding results at the destination, because the relay will not send a packet in the subsequent time slot after the joint state. Notice that from the Markov chain in Fig.~\ref{fig-NOMA-chain}, we see if the current state is state ${J^ - }$, the Markov chain may transit to state ${S^ + }$, ${S^ - }$, or ${S^ R }$ in the next time slot. However, if the current state is state ${J^ + }$, the Markov chain will transit to state ${S^ + }$ or ${S^ - }$ only, i.e., the system will not transit to ${S^ R }$. Recall that state ${S^ R }$ means that the destination successfully receives the old update packet from the relay but fails to decode the new update packet from the source in the NOMA transmission. Hence, if the current state is ${J^ - }$, the system transits to ${S^ R }$ with probability $p_{{\rm{rd}}}^{\rm{N}}(1 - {p_{{\rm{sd}}}}) = p_{{\rm{rd}}}^{\rm{N}} - p_{{\rm{sd}}}^{\rm{N}}$, meaning that the decoded packet forwarded by the relay leads to a successful update at the destination (i.e.,  the instantaneous AoI drops to two time slots). If the current state is ${J^ + }$, it means that the destination has already received the previous old packet. In the next time slot, the destination can exploit the received old packet to cancel the signals from the relay. Assuming perfect interference cancellation, the destination now tries to decode an interference-free packet from the source. 
If the new packet is decoded, the system will enter state ${S^ + }$; otherwise, the system will enter state ${S^ - }$. 

Let ${\Omega _{{\rm{NOMA}}}}$ denote the state transition matrix, which can be written by (\ref{f-matrix-NOMA}). The Markov chain in Fig.~\ref{fig-NOMA-chain} helps us derive the average AoI of the relay-assisted NOMA scheme.

\begin{figure*}[!t]
\setcounter{mytempeqncnt}{\value{equation}}
\footnotesize
\begin{align}
\resizebox{.99\hsize}{!}{%
${\Omega _{{\rm{NOMA}}}} =\begin{pmatrix}
    {{\omega _{{S^ + }{S^ + }}}} & {{\omega _{{S^ + }{J^ + }}}} & {{\omega _{{S^ + }{S^ R }}}} & {{\omega _{{S^ + }{J^ - }}}} & {{\omega _{{S^ + }{S^ - }}}} \cr 
    {{\omega _{{J^ + }{S^ + }}}} & {{\omega _{{J^ + }{J^ + }}}} & {{\omega _{{J^ + }{S^ R }}}} & {{\omega _{{J^ + }{J^ - }}}} & {{\omega _{{J^ + }{S^ - }}}} \cr 
    {{\omega _{{S^ R }{S^ + }}}} & {{\omega _{{S^ R }{J^ + }}}} & {{\omega _{{S^ R }{S^ R }}}} & {{\omega _{{S^ R }{J^ - }}}} & {{\omega _{{S^ R }{S^ - }}}} \cr 
    {{\omega _{{J^ - }{S^ + }}}} & {{\omega _{{J^ - }{J^ + }}}} & {{\omega _{{J^ - }{S^ R }}}} & {{\omega _{{J^ - }{J^ - }}}} & {{\omega _{{J^ - }{S^ - }}}} \cr  
    {{\omega _{{S^ - }{S^ + }}}} & {{\omega _{{S^ - }{J^ + }}}} & {{\omega _{{S^ - }{S^ R }}}} & {{\omega _{{S^ - }{J^ - }}}} & {{\omega _{{S^ - }{S^ - }}}} \cr  
\end{pmatrix} 
=\begin{pmatrix}
    {\left( {1 - {p_{{\rm{sr}}}}} \right){p_{{\rm{sd}}}}} & {{p_{{\rm{sr}}}}{p_{{\rm{sd}}}}} & 0 & {{p_{{\rm{sr}}}}\left( {1 - {p_{{\rm{sd}}}}} \right)} & {\left( {1 - {p_{{\rm{sr}}}}} \right)\left( {1 - {p_{{\rm{sd}}}}} \right)} \cr
    {p_{{\rm{sd}}}^{\rm{}}} & 0 & 0 & 0 & {1 - p_{{\rm{sd}}}^{\rm{}}} \cr
    {\left( {1 - {p_{{\rm{sr}}}}} \right){p_{{\rm{sd}}}}} & {{p_{{\rm{sr}}}}{p_{{\rm{sd}}}}} & 0 & {{p_{{\rm{sr}}}}\left( {1 - {p_{{\rm{sd}}}}} \right)} & {\left( {1 - {p_{{\rm{sr}}}}} \right)\left( {1 - {p_{{\rm{sd}}}}} \right)} \cr 
    {p_{{\rm{sd}}}^{\rm{N}}} & 0 & {p_{{\rm{rd}}}^{\rm{N}} - p_{{\rm{sd}}}^{\rm{N}}} & 0 & {1 - p_{{\rm{rd}}}^{\rm{N}}} \cr
    {\left( {1 - {p_{{\rm{sr}}}}} \right){p_{{\rm{sd}}}}} & {{p_{{\rm{sr}}}}{p_{{\rm{sd}}}}} & 0 & {{p_{{\rm{sr}}}}\left( {1 - {p_{{\rm{sd}}}}} \right)} & {\left( {1 - {p_{{\rm{sr}}}}} \right)\left( {1 - {p_{{\rm{sd}}}}} \right)} \cr  
\end{pmatrix}.$}
\label{f-matrix-NOMA}
\end{align}
\hrulefill
\end{figure*}

\textbf{Proposition 2:} The average AoI of the NOMA scheme, ${\bar \Delta _{{\rm{NOMA}}}}$, is computed by
\begin{align}
{\bar \Delta _{{\rm{NOMA}}}} &= \frac{{\mathbb{E}\left[ {\tau Z} \right]}}{{\mathbb{E}\left[ Z \right]}} + \frac{{\mathbb{E}\left[ {{Z^2}} \right]}}{{2\mathbb{E}\left[ Z \right]}} \notag \\
&= \frac{{1 + 3{p_{{\rm{sr}}}}}}{{2\left( {1 + {p_{{\rm{sr}}}}} \right)}} + \frac{{\left[ {1 + {p_{{\rm{sr}}}}\left( {1 - {p_{{\rm{sd}}}}} \right)} \right]\left[ {1 + {p_{{\rm{sr}}}}\left( {1 - p_{{\rm{sd}}}^{\rm{N}}} \right)} \right]}}{{\left[ {{p_{{\rm{sd}}}} + {p_{{\rm{sr}}}}p_{{\rm{rd}}}^{\rm{N}}\left( {1 - {p_{{\rm{sd}}}}} \right)} \right]\left( {1 + {p_{{\rm{sr}}}}} \right)}}  \notag \\
&~~~~+\frac{{\left[ {1 + {p_{{\rm{sr}}}}\left( {1 - {p_{{\rm{sd}}}}} \right)} \right]\left[ {{p_{{\rm{sr}}}}{p_{{\rm{sd}}}}\left( {p_{{\rm{sd}}}^{\rm{N}} - {p_{{\rm{sd}}}}} \right)} \right]}}{{\left[ {{p_{{\rm{sd}}}} + {p_{{\rm{sr}}}}p_{{\rm{rd}}}^{\rm{N}}\left( {1 - {p_{{\rm{sd}}}}} \right)} \right]\left( {1 + {p_{{\rm{sr}}}}} \right)}}.
\label{f-aoi-NOMA}
\end{align}

\begin{proof} See Appendix \ref{sec:appendix_B} for the details. The average AoI is computed following the same method as in the TDMA scheme using the Markov chain depicted in Fig. \ref{fig-NOMA-chain}.
\end{proof}

\section{Performance Evaluation }
\label{section-Performance} 
This section compares the average AoI among different schemes. We first estimate the packet error rate (PER) using the short packet theory. After that, we examine the average AoI of the non-relay and relay-assisted schemes (i.e., the TDMA and NOMA schemes). 

\subsection{Packet Error Rate for Short Packets }
\label{section-Error} 
We assume the channel of each wireless link in the status update system contains both small-scale block-fading and large-scale loss \cite{Tse2005,Zheng2021}. For each link, the SNR at the receiver is expressed by $\gamma  = \frac{{P{\theta    _0}\lvert h \rvert^2{d^{ - \varphi }}}}{{{\sigma ^2}}}$, where $P$ is the transmit power of the transmitter, ${{\theta_0}}$ is the channel power gain at the reference distance, $h$ is the small-scale block-fading Rayleigh distribution coefficient with zero mean and unit variance (i.e., $\mathbb{E}[\lvert h \rvert^2]=1$), $d$ is the distance between the transmitter and the receiver, $\varphi $ is the path loss exponent, and $\sigma^2 $ is the power of additive white Gaussian noise. Therefore,  the average SNR at the receiver is $\bar \gamma  = \frac{{P{\theta_0}{d^{ - \varphi }}}}{{{\sigma ^2}}}$. 

In status update systems, the update packets transmitted by the source are usually short. Information theory indicates that with finite block lengths, the PER cannot go to zero. Therefore, we use the short packet theory to estimate the PER. According to \cite{Polyanskiy2010}, the PER in AWGN channels can be expressed as
\begin{align}
\varepsilon  \approx Q\left( {\frac{{\sqrt n \left( {\log \left( {1 + \gamma } \right) - \frac{D}{n}} \right)}}{{\sqrt {1 - \frac{1}{{{{\left( {1 + \gamma } \right)}^2}}}} }}} \right),
\label{f-packet-error}
\end{align}
where $D$ and $n$ are the numbers of source bits and coded bits of an update packet, respectively, and ${\frac{D}{n}}$ is the code rate. $Q\left( \cdot \right)$ is the Q-function, i.e., $Q\left( x \right) = \frac{1}{{\sqrt {2\pi } }}\int_x^\infty  {{e^{ - \frac{{{t^2}}}{2}}}dt} $.

Under the quasi-static Rayleigh fading with an average SNR of $\bar\gamma$, we need to obtain the average PER. To do so, the Q-function in (\ref{f-packet-error}) can be approximated as \cite{Zheng2021}
\begin{align}
\varepsilon  \approx \left\{ 
\begin{aligned}
&1&,\; & \gamma  < \delta  + \frac{1}{{2\beta }},\\
&\beta \left( {\gamma  - \delta } \right) + \frac{1}{2}&,\; & \delta  + \frac{1}{{2\beta }} \le \gamma  \le \delta  - \frac{1}{{2\beta }}\\
&0&,\; & \gamma  > \delta  - \frac{1}{{2\beta }},
\end{aligned} \right.,
\label{f-Q-function}
\end{align}
where $\delta  = 2^{ {\frac{D}{n}}} - 1$ and $\beta  =  - \sqrt {\frac{n}{{2\pi \left( {2^ {\left( {\frac{2D}{n}} \right)} - 1} \right)}}} $. Therefore, the average PER ${\bar \varepsilon }$ can be computed by  (\ref{f-average-packet-error}). Based on the PER formula (\ref{f-average-packet-error}), we can estimate ${p_{{\rm{sd}}}}$, ${p_{{\rm{sr}}}}$, and ${p_{{\rm{rd}}}}$ (i.e., one minus the corresponding PER) under different $\bar \gamma$, which are used to evaluate the average AoI presented in the next subsection. For the NOMA scheme, $p_{{\rm{rd}}}^{\rm{N}}$ and $p_{{\rm{sd}}}^{\rm{N}}$ are computed by (\ref{f-p_rd_noma}) and (\ref{f-p_sd_noma}), respectively, where $Ei(\cdot)$ denotes the exponential integral (see Appendix \ref{sec:appendix_C} for the detailed computation).

\begin{figure*}[!t]
\setcounter{mytempeqncnt}{\value{equation}}
\footnotesize
\begin{align}
\bar \varepsilon  &= \int_0^\infty  {{f_\gamma }\left( x \right)Q\left( {\frac{{\sqrt n \left( {{{\log }}\left( {1 + \gamma } \right) - \frac{D}{n}} \right)}}{{\sqrt {1 - \frac{1}{{{{\left( {1 + \gamma } \right)}^2}}}} }}} \right)dx} =\int_0^{\delta  + \frac{1}{{2\beta }}} {{f_\gamma }\left( x \right)dx}  + \int_{\delta  + \frac{1}{{2\beta }}}^{\delta  - \frac{1}{{2\beta }}} {{f_\gamma }\left( x \right)\left[ {\beta \left( {x - \delta } \right) + \frac{1}{2}} \right]dx}  \notag \\
&= \int_0^{\delta  + \frac{1}{{2\beta }}} {{f_\gamma }\left( x \right)dx}  + \left( {\frac{1}{2} - \delta \beta } \right)\int_{\delta  + \frac{1}{{2\beta }}}^{\delta  - \frac{1}{{2\beta }}} {{f_\gamma }\left( x \right)dx}  + \beta \int_{\delta  + \frac{1}{{2\beta }}}^{\delta  - \frac{1}{{2\beta }}} {x{f_\gamma }\left( x \right)dx} 
= 1 + \beta \bar \gamma \left( {{{\rm{e}}^{ - \frac{1}{{\bar \gamma }}\left( {\delta  + \frac{1}{{2\beta }}} \right)}} - {{\rm{e}}^{ - \frac{1}{{\bar \gamma }}\left( {\delta  - \frac{1}{{2\beta }}} \right)}}} \right).
\label{f-average-packet-error}
\end{align}
\hrulefill
\end{figure*}

\begin{figure*}[!t]
\setcounter{mytempeqncnt}{\value{equation}}
\footnotesize
\begin{align}
p_{{\rm{rd}}}^{\rm{N}} &=  \beta \frac{{{{\bar \gamma }_{{\rm{rd}}}}}}{{{{\bar \gamma }_{{\rm{sd}}}}}}{e^{\frac{1}{{{{\bar \gamma }_{{\rm{sd}}}}}}}}\left( {Ei\left( { - \left( {\frac{1}{{{{\bar \gamma }_{{\rm{sd}}}}}} + \frac{{\delta  + \frac{1}{{2\beta }}}}{{{{\bar \gamma }_{{\rm{rd}}}}}}} \right)} \right) - Ei\left( { - \left( {\frac{1}{{{{\bar \gamma }_{{\rm{sd}}}}}} + \frac{{\delta  - \frac{1}{{2\beta }}}}{{{{\bar \gamma }_{{\rm{rd}}}}}}} \right)} \right)} \right). \label{f-p_rd_noma}\\
p_{{\rm{sd}}}^{\rm{N}} &= p_{{\rm{rd}}}^{\rm{N}}{p_{{\rm{sd}}}} = {\beta ^2}{\bar \gamma _{{\rm{rd}}}}{e^{\frac{1}{{{{\bar \gamma }_{{\rm{sd}}}}}}}}\left( {{e^{ - \frac{1}{{{{\bar \gamma }_{{\rm{sd}}}}}}\left( {\delta  - \frac{1}{{2\beta }}} \right)}} - {e^{ - \frac{1}{{{{\bar \gamma }_{{\rm{sd}}}}}}\left( {\delta  + \frac{1}{{2\beta }}} \right)}}} \right)\left( {Ei\left( { - \left( {\frac{1}{{{{\bar \gamma }_{{\rm{sd}}}}}} + \frac{{\delta  + \frac{1}{{2\beta }}}}{{{{\bar \gamma }_{{\rm{rd}}}}}}} \right)} \right) - Ei\left( { - \left( {\frac{1}{{{{\bar \gamma }_{{\rm{sd}}}}}} + \frac{{\delta  - \frac{1}{{2\beta }}}}{{{{\bar \gamma }_{{\rm{rd}}}}}}} \right)} \right)} \right). \label{f-p_sd_noma}
\end{align}
\hrulefill
\end{figure*}

\subsection{Average AoI versus Code Rate}
\label{section-aoI-code tate} 
\begin{figure}
\centerline{\includegraphics[width=0.42\textwidth]{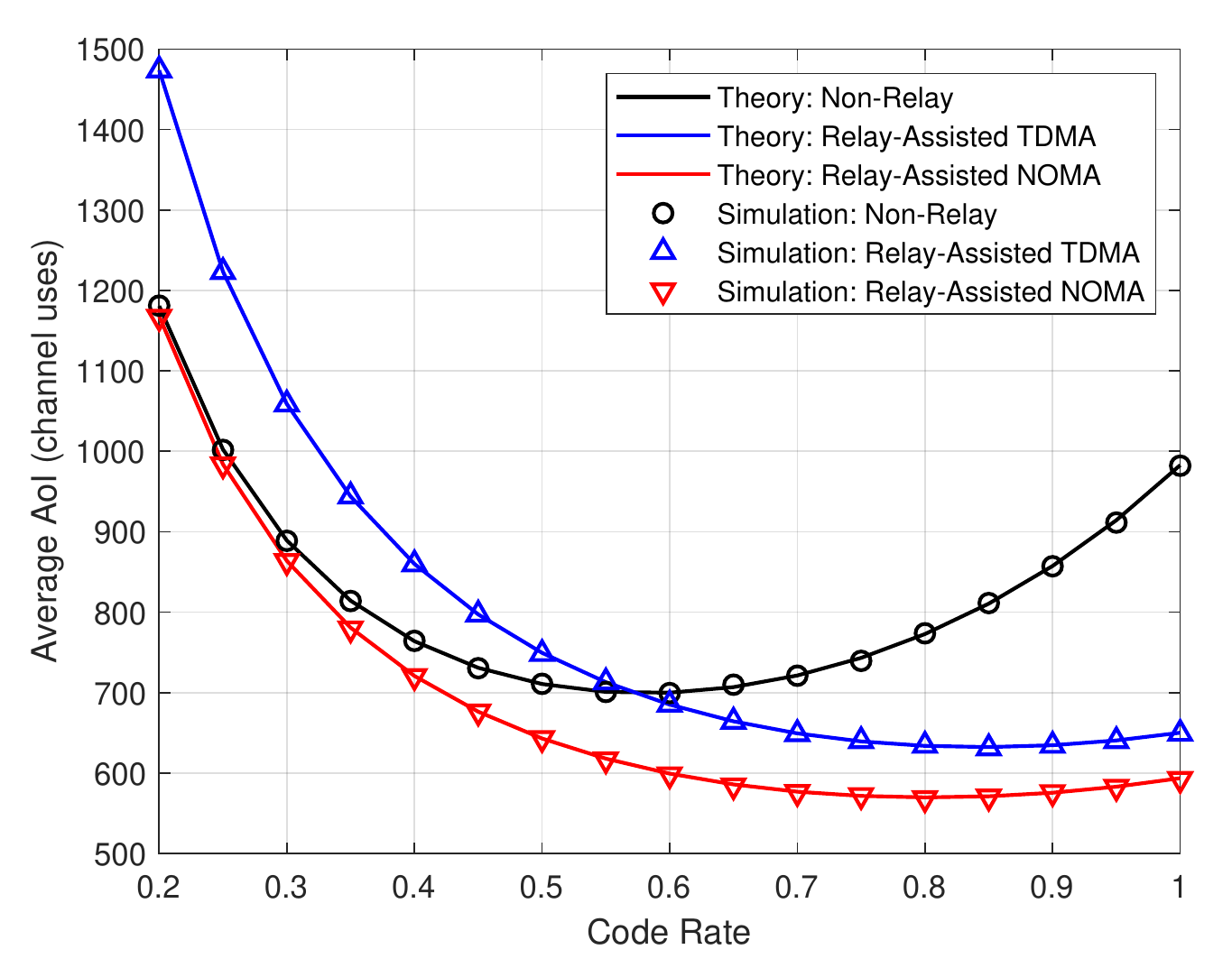}}
  \caption{Performance comparison for the non-relay scheme and the relay-assisted TDMA and NOMA schemes: the average AoI versus the code rate, where ${{\bar\gamma _{{\rm{sd}}}}} = -3dB$, ${\bar\gamma _{{\rm{rd}}}} = {\bar\gamma _{{\rm{sr}}}} = 3dB$. The payload of an update packet is 128 bits.}\label{fig-diff-code rate}
\end{figure}

We first examine the relationship between the average AoI and the code rate. Fig.~\ref{fig-diff-code rate} plots the average AoI versus the code rate for different schemes, namely the non-relay scheme, the relay-assisted TDMA scheme, and the relay-assisted NOMA scheme. The number of source bits per update packet is $128$ bits. In this simulation, the average SNR from the source to the destination ${{\bar\gamma _{{\rm{sd}}}}}$ is assumed to be $-3dB$.  The average SNR of the relay at the destination and the average SNR of the source at the relay are both set to $3 dB$, i.e., ${{\bar\gamma _{{\rm{rd}}}}}={{\bar\gamma _{{\rm{sr}}}}}=3dB$. 

We plot both the theoretical and simulation results in  Fig.~\ref{fig-diff-code rate} to verify the theoretical analysis of the average AoI derived in the previous sections. Specifically, for the theoretical results, we first estimate the PERs and then substitute them to the corresponding average AoI formulas, i.e., we substitute ${p_{{\rm{sd}}}}$ into (\ref{f-AoI-NR}) for the non-relay scheme, substitute ${p_{{\rm{sd}}}}$, ${p_{{\rm{sr}}}}$ and ${p_{{\rm{rd}}}}$ into (\ref{f-aoi-OMA}) for the relay-assisted TDMA scheme, and substitute ${p_{{\rm{sd}}}}$, ${p_{{\rm{sr}}}}$, $p_{{\rm{rd}}}^{\rm{N}}$, and $p_{{\rm{sd}}}^{\rm{N}}$ into (\ref{f-aoi-NOMA}) for the relay-assisted NOMA scheme. For the simulation results, we first conduct simulations of the three protocols based on ${p_{{\rm{sd}}}}$, ${p_{{\rm{sr}}}}$, ${p_{{\rm{rd}}}}$, $p_{{\rm{rd}}}^{\rm{N}}$ and $p_{{\rm{sd}}}^{\rm{N}}$ over a large number of time slots. We collect the instantaneous AoI in each time slot based on the decoding results, from which the average AoI is calculated. Also, notice that the unit of the average AoI in Fig.~\ref{fig-diff-code rate} is the number of channel uses, which is simply the multiplication of the blocklength $n$ of an update packet and the average AoI in the number of time slots.

From Fig.~\ref{fig-diff-code rate}, we see that the simulation results corroborate the theoretical results, thus validating the correctness of our derivations in the average AoI of different schemes. In particular, the relay-assisted NOMA scheme outperforms both the non-relay and relay-assisted TDMA schemes in terms of the average AoI, when the channel condition from the source to the destination is weak (i.e., ${{\bar\gamma _{{\rm{sd}}}}}=-3dB$). The detailed average AoI comparison among different schemes will be presented in Section \ref{section-aoI-comparison}.

Interestingly, as indicated by Fig.~\ref{fig-diff-code rate}, all the three schemes can achieve their optimal average AoI when the code rate is neither too high nor too low. The reasons are as follows. Given a fixed number of source bits, when the code rate is low, the number of coded bits is large (i.e., more coding redundancy is added), thus resulting in a longer duration for each update packet. Although adding more coding redundancy reduces the PER, the long packet duration leads to a high average AoI. In contrast, when the code rate is high, the duration of an update packet becomes shorter at the cost of higher PERs. Thus, the destination suffers from high PERs and receives fewer update packets, leading to a high average AoI as well. Therefore, as the code rate increases, the average AoI first decreases and then increases. As shown in Fig.~\ref{fig-diff-code rate}, there exist optimal code rates to achieve the lowest average AoI for different schemes.

\begin{figure}
\centerline{ \includegraphics[width=0.4\textwidth]{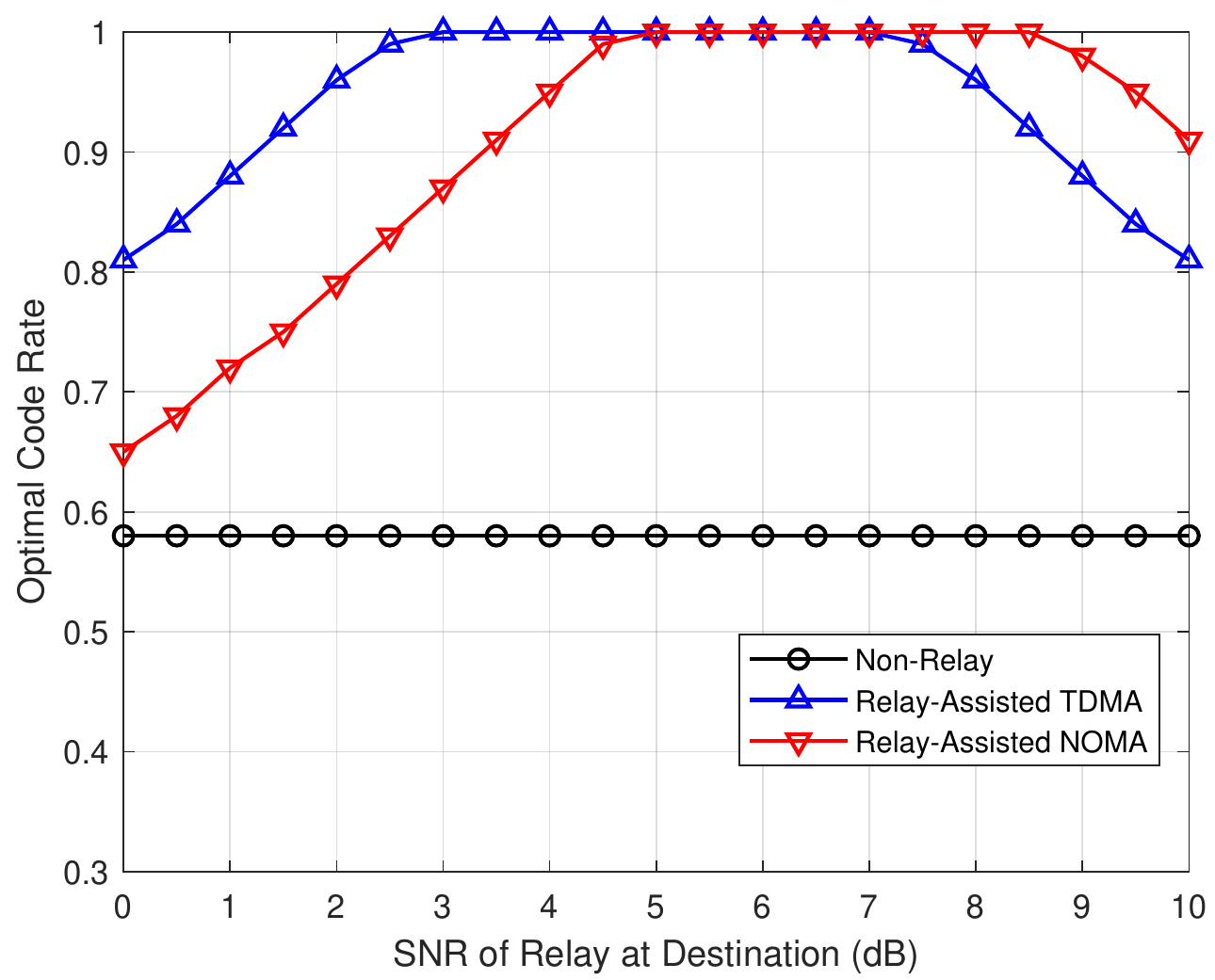}}
  \caption{The optimal code rate versus the average SNR from the relay at the destination ${\bar\gamma _{{\rm{rd}}}}$, where ${{\bar\gamma _{{\rm{sd}}}}} = -3dB$, ${\bar\gamma _{{\rm{rd}}}} + {\bar\gamma _{{\rm{sr}}}} = 10dB$. The payload of an update packet is 128 bits.}\label{fig-optimal-code rate}
\end{figure}

In general, we need to optimize the code rate for different SNR scenarios to achieve the corresponding optimal average AoI. To see this, Fig. \ref{fig-optimal-code rate} plots the optimal code rate versus the average SNR from the relay at the destination ${\bar\gamma _{{\rm{rd}}}}$. We consider the average SNR from the source to the destination ${{\bar\gamma _{{\rm{sd}}}}} = -3dB$ as an example, and other ${{\bar\gamma _{{\rm{sd}}}}}$ leads to similar observations. In addition, ${\bar\gamma _{{\rm{rd}}}} + {\bar\gamma _{{\rm{sr}}}} = 10dB$. That is, when ${{\bar\gamma _{{\rm{rd}}}}}$ increases from $0dB$ to $10dB$, the average SNR of the source at the relay ${{\bar\gamma _{{\rm{sr}}}}}$ decreases from $10dB$ to $0dB$ at the same time. This SNR setting simulates the impact of different locations of the relay. 

First, we see from Fig. \ref{fig-optimal-code rate} that the optimal code rate for the non-relay scheme is a constant of around 0.58 under different ${\bar\gamma _{{\rm{rd}}}}$, since there is only a direct link with a fixed SNR ${{\bar\gamma _{{\rm{sd}}}}} = -3dB$. Next, we observe that as ${\bar\gamma _{{\rm{rd}}}}$ increases, the optimal code rate for the relay-assisted TDMA scheme first increases and then decreases. As indicated by Proposition 1, i.e., equation (\ref{f-aoi-OMA}), when ${\bar\gamma _{{\rm{sd}}}}$ (and also ${{p_{{\rm{sd}}}}}$) is fixed, the average AoI of the relay-assisted TDMA scheme ${{\bar \Delta }_{{\rm{TDMA}}}}$ is inversely proportional to the product of ${{p_{{\rm{sr}}}}}$ and ${{p_{{\rm{rd}}}}}$. When the channel condition of any hop is low (i.e., either ${\bar\gamma _{{\rm{rd}}}}$ or ${\bar\gamma _{{\rm{sr}}}}$ is low), it requires a lower code rate to achieve a good overall PER performance at the destination. Therefore, we see from Fig. \ref{fig-optimal-code rate} that the optimal code rate for the TDMA scheme is symmetric with respect to ${\bar\gamma _{{\rm{rd}}}} = {\bar\gamma _{{\rm{sr}}}} = 5dB$. In other words, when both hops have the same channel condition and have a relatively high SNR$=5dB$, a large code rate should be chosen to shorten the duration of update packets, thereby lowering the average AoI.

Similar to the relay-assisted TDMA scheme, as the average SNR ${\bar\gamma _{{\rm{rd}}}}$ increases, Fig. \ref{fig-optimal-code rate} shows that the optimal code rate for the relay-assisted NOMA scheme also first increases and then decreases. Recall that in the NOMA scheme, when the destination receives the superimposed packets from the relay and the source, the SIC decoder first decodes the packet from the relay by treating the packet from the source as noise. Due to SIC, the effective SNR from the relay to the destination becomes lower. If ${\bar\gamma _{{\rm{rd}}}}$ is as small as $0dB$, more coding redundancy should be added to have a better PER performance at the second hop, i.e., a lower code rate is used. When ${\bar\gamma _{{\rm{rd}}}}$ becomes larger (e.g., $6dB$), a larger code rate should be chosen to shorten the duration of update packets and lower the average AoI. When ${\bar\gamma _{{\rm{rd}}}}$ is as large as $10dB$ such that ${\bar\gamma _{{\rm{sr}}}}$ becomes $0dB$, the optimal code rate now should be reduced to ensure that the relay can receive update packets from the source in the first hop first, before forwarding to the destination in the second hop. We notice that a lower code rate is needed when ${\bar\gamma _{{\rm{rd}}}}$ is $0dB$ (more coding redundancy), compared with the scenario when ${\bar\gamma _{{\rm{rd}}}}$ is $10dB$ (less coding redundancy). This indicates that the PER performance caused by the SIC operation in the second hop affects the optimal average AoI more significantly than the PER in the first hop. In the next subsection, we examine the optimal average AoI given the optimal code rates under different SNR scenarios. 

\subsection{Average AoI Comparison Among Different Schemes}\label{section-aoI-comparison}
\begin{figure*}
\centerline{\includegraphics[width=0.99\linewidth]{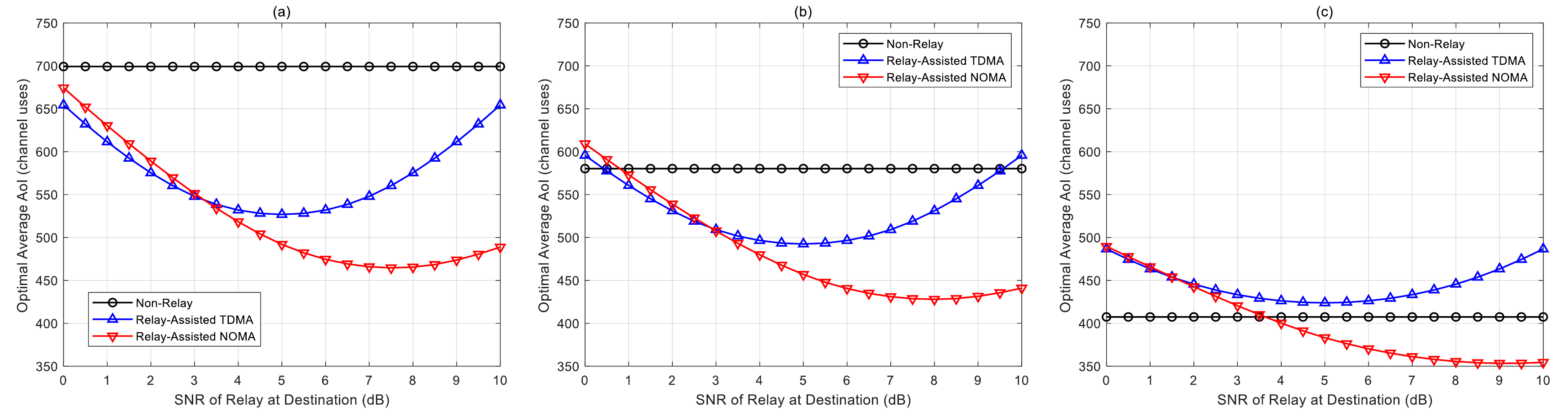}}
\caption{Performance comparison for the non-relay scheme and the relay-assisted TDMA and NOMA schemes: the optimal average AoI versus ${{\bar\gamma _{{\rm{rd}}}}}$ with ${\bar\gamma _{{\rm{sr}}}} + {\bar\gamma _{{\rm{rd}}}} = 10dB$, when ${{\bar\gamma _{{\rm{sd}}}}}$ is (a) $-3 dB$, (b) $-2 dB$, and (c) $0 dB$. The payload of an update packet is 128 bits.}
\label{fig-diff-SNR}
\end{figure*}

\emph{\underline{Answers to \textbf{Q1} in Section I:}} We now compare the average AoI of the non-relay and the relay-assisted schemes. Specifically, Figs.~\ref{fig-diff-SNR}(a)--(c) plot the average AoI versus the average SNR of the relay at the destination ${{\bar\gamma _{{\rm{rd}}}}}$, when the average SNR of the source at the destination ${{\bar\gamma _{{\rm{sd}}}}}$ is $-3 dB$, $-2 dB$, and $0 dB$, respectively. As shown in Fig.~\ref{fig-diff-SNR}, compared with the non-relay scheme, the relay-assisted schemes improve the information freshness greatly when the ${\bar\gamma _{{\rm{sd}}}}$ is weak. For example, when ${\bar\gamma _{{\rm{sd}}}}=-3dB$, both the relay-assisted TDMA and NOMA schemes outperform the non-relay scheme by giving a significantly lower average AoI. In contrast, when ${\bar\gamma _{{\rm{sd}}}}=0dB$, the non-relay scheme generally gives a lower average AoI. This shows that the dedicated relay does not always help reduce the average AoI when the destination can receive an update packet via the direct link under a certain successful probability (e.g., when ${\bar\gamma _{{\rm{sd}}}}=0dB$, the successful reception probability is around $0.4$). Such results reveal the key difference when a direct link is involved in relay-assisted networks compared with prior works focusing on the relay-assisted link only  \cite{Feng2022,Xie2021relay, Moradian2020,Zheng2021}. 

When ${\bar\gamma _{{\rm{sd}}}}=-2dB$, as shown in Fig.~\ref{fig-diff-SNR}(b), the average AoI of the relay-assisted TDMA scheme is still lower than that of the non-relay scheme in most scenarios, i.e., when the SNR of any hop is neither too low nor too high. Only when ${\bar\gamma_{{\rm{rd}}}}$ or ${\bar\gamma_{{\rm{sr}}}}$ is $0dB$, the average AoI of the non-relay scheme outperforms that of the TDMA scheme. This shows that when one of the two hops suffers from a very poor channel condition, a dedicated relay does not help improve information freshness if TDMA is used. In contrast, by using the NOMA scheme, the average AoI is lower than that of the non-relay scheme as long as ${\bar\gamma _{{\rm{rd}}}}$ is larger than $1dB$, as plotted in Fig.~\ref{fig-diff-SNR}(b).

\emph{\underline{Answers to \textbf{Q2} in Section I:}} Let us now focus more on the two relay-assisted schemes. Figs.~\ref{fig-diff-SNR}(a) and \ref{fig-diff-SNR}(b) show that the average AoI of the relay-assisted TDMA scheme first decreases when ${\bar\gamma _{{\rm{rd}}}}$ increases from $0 dB$ to $5dB$, and then increases when ${\bar\gamma _{{\rm{rd}}}}$ continues increasing to $10dB$. In other words, the lowest average AoI of the relay-assisted TDMA scheme is achieved when ${\bar \gamma _{{\rm{sr}}}} = {\bar \gamma _{{\rm{rd}}}} = 5dB$. Since the average AoI ${{\bar \Delta }_{{\rm{TDMA}}}}$ is inversely proportional to the product of ${{p_{{\rm{sr}}}}}$ and ${{p_{{\rm{rd}}}}}$, the optimal average AoI achieves the lowest when both hops have a relatively high SNR of $5dB$. Similar to OMA, the average AoI of the NOMA scheme also first decreases and then increases. However, as indicated in Fig.~\ref{fig-diff-SNR}(a) (and Fig. \ref{fig-diff-SNR}(b)), unlike the TDMA scheme in which the lowest average AoI is achieved when ${\bar\gamma _{{\rm{rd}}}}=5dB$, the NOMA scheme achieves the lowest average AoI when ${\bar\gamma_{{\rm{rd}}}}=7.5dB$ (and ${\bar\gamma_{{\rm{rd}}}}=8dB$). This phenomenon also indicates that the PER performance in the second hop affects the optimal average AoI more significantly than the one in the first hop. 

Comparing the average AoI between the NOMA and TDMA schemes, we see from Figs.~\ref{fig-diff-SNR}(a) and \ref{fig-diff-SNR}(b)) that when ${\bar\gamma_{{\rm{rd}}}}$ is small, the TDMA scheme gives a lower average AoI than the NOMA scheme. This is because the effective SNR from the relay to the destination in the NOMA scheme is small, so the successful decoding probability by SIC is low at the destination. When ${\bar\gamma_{{\rm{rd}}}}$ increases, e.g., ${\bar\gamma_{{\rm{rd}}}}\geq 4dB$ in Fig.~\ref{fig-diff-SNR}(a), the relay-assisted NOMA scheme now outperforms its TDMA counterpart. This performance improvement is mainly due to the higher packet generation and transmission rate at the source in the NOMA scheme. Recall that in the TDMA scheme, the source sends a new update packet every other time slot. In the NOMA scheme, it sends a new update packet in every time slot so that the destination has a chance to receive the most up-to-date update packet in every time slot. For example, suppose the source sends an update packet, and the relay fails to decode the update packet from the source in the current time slot. In that case, the TDMA scheme will waste the next time slot because the source remains silent to avoid possible interference with the relay at the destination. Note that since there is no feedback, the source does not know the decoding result of the relay. However, the destination in the NOMA scheme will send a new update packet in the next time slot, regardless of the decoding outcome of the relay, so that the destination can have the opportunity to update its status immediately (i.e., no time slot will be idle). This saves the airtime and reduces the average AoI. Furthermore, since the NOMA scheme depends much on the success of SIC, we observe that its average AoI is sensitive to ${\bar\gamma_{{\rm{rd}}}}$, i.e., the SNR of the second hop. For example, in Figs.~\ref{fig-diff-SNR}(a) and \ref{fig-diff-SNR}(b), whether NOMA outperforms TDMA depends greatly on ${\bar\gamma_{{\rm{rd}}}}$. In contrast, TDMA scheme has a more stable average AoI across different ${\bar\gamma_{{\rm{rd}}}}$. This property is beneficial when the channel conditions vary quickly in practical wireless systems. 

\begin{figure}
\centerline{ \includegraphics[width=0.38\textwidth]{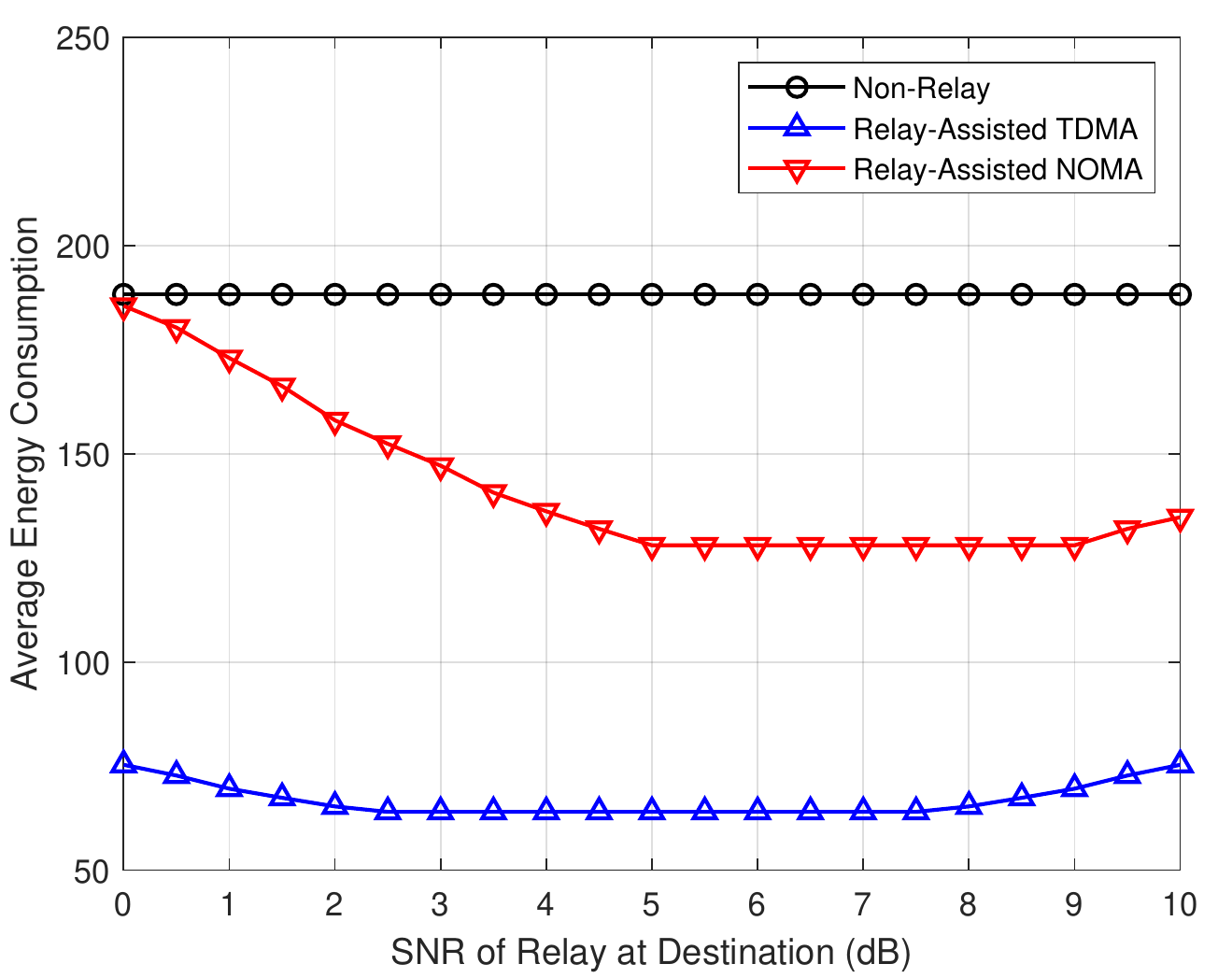}}
\caption{Performance comparison for the non-relay scheme and the relay-assisted TDMA and NOMA schemes: the average energy consumption versus ${{\bar\gamma _{{\rm{rd}}}}}$, when ${{\bar\gamma _{{\rm{sd}}}}}=-2dB$. The payload of an update packet is 128 bits.}
\label{fig-energy}
\end{figure}

\emph{\underline{Discussion on the Energy Issue:}} At a higher packet generation and transmission rate, it is easy to see that the source in the NOMA scheme consumes more energy than that in the TDMA scheme. Fig.~\ref{fig-energy} plots the average energy consumption of the source versus ${\bar\gamma _{{\rm{rd}}}}$, when ${\bar\gamma_{{\rm{sd}}}}=-2dB$. Suppose that the energy to transmit a coded bit is $E=1$. Then the average energy consumption $\bar E$ is defined as the energy to transmit the whole update packet consisting of $n$ coded bits under the optimal code rate, multiplied by the fraction of time slots $V$ that the source will send an update packet, i.e.,  $\bar E = nEV$. Note that $V=1$ for the non-relay and relay-assisted NOMA schemes, and $V=0.5$ for the relay-assisted TDMA scheme. Fig.~\ref{fig-energy} shows that the relay-assisted TDMA scheme offers the lowest energy consumption. Hence, the TDMA scheme is a more viable option for energy-efficient relay-assisted status update systems when the information freshness requirement is not very stringent. On the other hand, if the energy issue is not a concern for the source, i.e., a higher packet generation and transmission rate is possible, the NOMA scheme should be adopted when the relay has a high enough SNR at the destination. Otherwise, the TDMA scheme is a more suitable solution since it provides a more stable average AoI across different channel conditions. 

\section{Conclusions}
We have investigated the average AoI of a relay-assisted system where a source node wants to send status update packets to a destination node as timely as possible with the help of a relay node. Specifically, we theoretically study the average AoI of two relay-assisted schemes, namely the TDMA and NOMA schemes, and examine their merits under different system settings.  

This paper aims to answer the question (i.e., \textbf{Q1}): whether the dedicated relay can improve the AoI performance of a relay-assisted system with a direct link, compared with a non-relay system. The exploration of \textbf{Q1} leads to another interesting question (i.e., \textbf{Q2}): \emph{what packet generation and transmission strategy should be adopted by the source when a dedicated relay is involved in the system?} To this end, we investigate relay-assisted TDMA and NOMA schemes, depending on whether the source and the relay are allowed to send simultaneously. A key challenge in deriving the theoretical average AoI is that the destination has different successful probabilities for receiving an update packet in different time slots. To tackle this issue, we establish Markov chains to model different transmission schemes, from which the average AoI can be derived.  

Both theoretical and simulation results indicate that due to the existence of the direct link, the relay-assisted schemes do not always outperform the non-relay scheme (i.e., the answer to \textbf{Q1}). When the SNR of the source-destination link is as weak as $-2dB$, both relay-assisted TDMA and NOMA schemes significantly outperform the non-relay scheme in terms of average AoI. For \textbf{Q2}, we show that if the SNR of the relay-destination link is sufficiently high compared with the source-destination link such that SIC works well, the source should generate and send an update packet every time slot. In other words, the NOMA scheme should be used since it reduces the average AoI by as much as 25\% compared to its TDMA counterpart. When the energy issue is a concern for the source and the AoI requirement is less stringent, the TDMA scheme, in which the source generates and sends an update packet every other time slot, is an effective solution for low-power relay-assisted status update systems.

\appendices
\section{The Average AoI of the TDMA Relay-Assisted Scheme with Maximum Ratio Combining (MRC) } \label{sec:appendix_A}
We now investigate the theoretical average AoI of the relay-assisted TDMA scheme when maximum ratio combining (MRC) is used at destination. Using MRC increases the decoding probability at the destination, which is now denoted by $p_{{\rm{rd}}}^{{\rm{MRC}}}$. By Proposition 1, the average AoI of the relay-assisted TDMA transmission scheme, $\bar \Delta _{{\rm{TDMA}}}^{{\rm{MRC}}}$, is given by replacing $p_{{\rm{rd}}}$ with $p_{{\rm{rd}}}^{{\rm{MRC}}}$ in (\ref{f-aoi-OMA}).

The probability density function (PDF) of the SNR ${\gamma _{{\rm{sd}}}}$ and ${\gamma _{{\rm{rd}}}}$ is
\begin{align}
{f_{{\gamma _v}}}\left( x \right) = \frac{{\exp \left( { - x/{{\bar \gamma }_v}} \right)}}{{{{\bar \gamma }_v}}},v \in \{ {\rm{sd}},{\rm{rd}}\} 
\end{align}
When the destination uses MRC to combine the received signals from the relay and the source to jointly decode an update packet, the cumulative distribution function (CDF) of the combined SNR ${\gamma ^{{\rm{MRC}}}}$ is
\begin{align}
&{F_{{\gamma ^{{\rm{MRC}}}}}}\left( x \right) = \Pr \left( {{\gamma _{{\rm{sd}}}} + {\gamma _{{\rm{rd}}}} \le x} \right) \notag \\
&= \int_0^x {P\left( {{\gamma _{{\rm{sd}}}} + {\gamma _{{\rm{rd}}}} \le x\left| {{\gamma _{{\rm{sd}}}} = {x_1}} \right.} \right)\frac{{\exp \left( { - {x_1}/{{\bar \gamma }_{{\rm{sd}}}}} \right)}}{{{{\bar \gamma }_{{\rm{sd}}}}}}d{x_1}} \notag \\
& = \int_0^x {P\left( {{\gamma _{{\rm{rd}}}} \le x - {x_1}} \right)\frac{{\exp \left( { - {x_1}/{{\bar \gamma }_{{\rm{sd}}}}} \right)}}{{{{\bar \gamma }_{{\rm{sd}}}}}}d{x_1}} \notag \\
& = \int_0^x {\left( {1 - \exp \left( { - \left( {x - {x_1}} \right)/{{\bar \gamma }_{{\rm{rd}}}}} \right)} \right)\frac{{\exp \left( { - {x_1}/{{\bar \gamma }_{{\rm{sd}}}}} \right)}}{{{{\bar \gamma }_{{\rm{sd}}}}}}d{x_1}} \notag \\
& = 1 - \exp \left( { - x/{{\bar \gamma }_{{\rm{sd}}}}} \right) + \frac{{{{\bar \gamma }_{{\rm{rd}}}}}}{{{{\bar \gamma }_{{\rm{rd}}}} - {{\bar \gamma }_{{\rm{sd}}}}}}\left( {\exp \left( { - x/{{\bar \gamma }_{{\rm{sd}}}}} \right) - \exp \left( { - x/{{\bar \gamma }_{{\rm{rd}}}}} \right)} \right)\notag \\
& = 1 + \frac{{{{\bar \gamma }_{{\rm{sd}}}}}}{{{{\bar \gamma }_{{\rm{rd}}}} - {{\bar \gamma }_{{\rm{sd}}}}}}\exp \left( { - x/{{\bar \gamma }_{{\rm{sd}}}}} \right) - \frac{{{{\bar \gamma }_{{\rm{rd}}}}}}{{{{\bar \gamma }_{{\rm{rd}}}} - {{\bar \gamma }_{{\rm{sd}}}}}}\exp \left( { - x/{{\bar \gamma }_{{\rm{rd}}}}} \right).
\end{align}
Therefore, the PDF of ${\gamma ^{{\rm{MRC}}}}$ is 
\begin{align}
{f_{{\gamma ^{{\rm{MRC}}}}}}\left( x \right) = \frac{{\exp \left( { - x/{{\bar \gamma }_{{\rm{rd}}}}} \right) - \exp \left( { - x/{{\bar \gamma }_{{\rm{sd}}}}} \right)}}{{{{\bar \gamma }_{{\rm{rd}}}} - {{\bar \gamma }_{{\rm{sd}}}}}}
\end{align}
Now $\bar \Delta _{{\rm{TDMA}}}^{{\rm{MRC}}}$ can be computed by $p_{rd}^{{\rm{MRC}}} = 1 - {\bar \varepsilon ^{{\rm{MRC}}}}$, where ${\bar \varepsilon ^{{\rm{MRC}}}}$ is the PER at the destination when MRC is used, which is computed by (\ref{f-MRC-equation}). 

\begin{figure*}[!t]
\setcounter{mytempeqncnt}{\value{equation}}
\footnotesize
\begin{align}
{{\bar \varepsilon }^{{\rm{MRC}}}} &= \int_0^\infty  {\frac{{\exp \left( { - x/{{\bar \gamma }_{{\rm{rd}}}}} \right) - \exp \left( { - x/{{\bar \gamma }_{{\rm{sd}}}}} \right)}}{{{{\bar \gamma }_{{\rm{rd}}}} - {{\bar \gamma }_{{\rm{sd}}}}}}Q\left( {\frac{{\sqrt n \left( {\log \left( {1 + x} \right) - \frac{D}{n}} \right)}}{{\sqrt {1 - \frac{1}{{{{\left( {1 + x} \right)}^2}}}} }}} \right)dx} \notag \\
 &= \int_0^{\delta  + \frac{1}{{2\beta }}} {\frac{1}{{{{\bar \gamma }_{{\rm{rd}}}} - {{\bar \gamma }_{{\rm{sd}}}}}}\left( {\exp \left( { - \frac{x}{{{{\bar \gamma }_{{\rm{rd}}}}}}} \right) - \exp \left( { - \frac{x}{{{{\bar \gamma }_{{\rm{sd}}}}}}} \right)} \right)dx}  + \int_{\delta  + \frac{1}{{2\beta }}}^{\delta  - \frac{1}{{2\beta }}} {\frac{1}{{{{\bar \gamma }_{{\rm{rd}}}} - {{\bar \gamma }_{{\rm{sd}}}}}}\left( {\exp \left( { - \frac{x}{{{{\bar \gamma }_{{\rm{rd}}}}}}} \right) - \exp \left( { - \frac{x}{{{{\bar \gamma }_{{\rm{sd}}}}}}} \right)} \right)\left[ {\beta \left( {x - \delta } \right) + \frac{1}{2}} \right]dx} \notag \\
 &= \frac{\beta }{{{{\bar \gamma }_{{\rm{rd}}}} - {{\bar \gamma }_{{\rm{sd}}}}}}\left( {\left( {\exp \left( { - \frac{{\delta  + \frac{1}{{2\beta }}}}{{{{\bar \gamma }_{{\rm{rd}}}}}}} \right) - \exp \left( { - \frac{{\delta  - \frac{1}{{2\beta }}}}{{{{\bar \gamma }_{{\rm{rd}}}}}}} \right)} \right){{\left( {{{\bar \gamma }_{{\rm{rd}}}}} \right)}^2} - \left( {\exp \left( { - \frac{{\delta  + \frac{1}{{2\beta }}}}{{{{\bar \gamma }_{{\rm{sd}}}}}}} \right) - \exp \left( { - \frac{{\delta  - \frac{1}{{2\beta }}}}{{{{\bar \gamma }_{{\rm{sd}}}}}}} \right)} \right){{\left( {{{\bar \gamma }_{{\rm{sd}}}}} \right)}^2}} \right)
\label{f-MRC-equation}
\end{align}
\hrulefill
\end{figure*}

\begin{figure}
\centerline{ \includegraphics[width=0.42\textwidth]{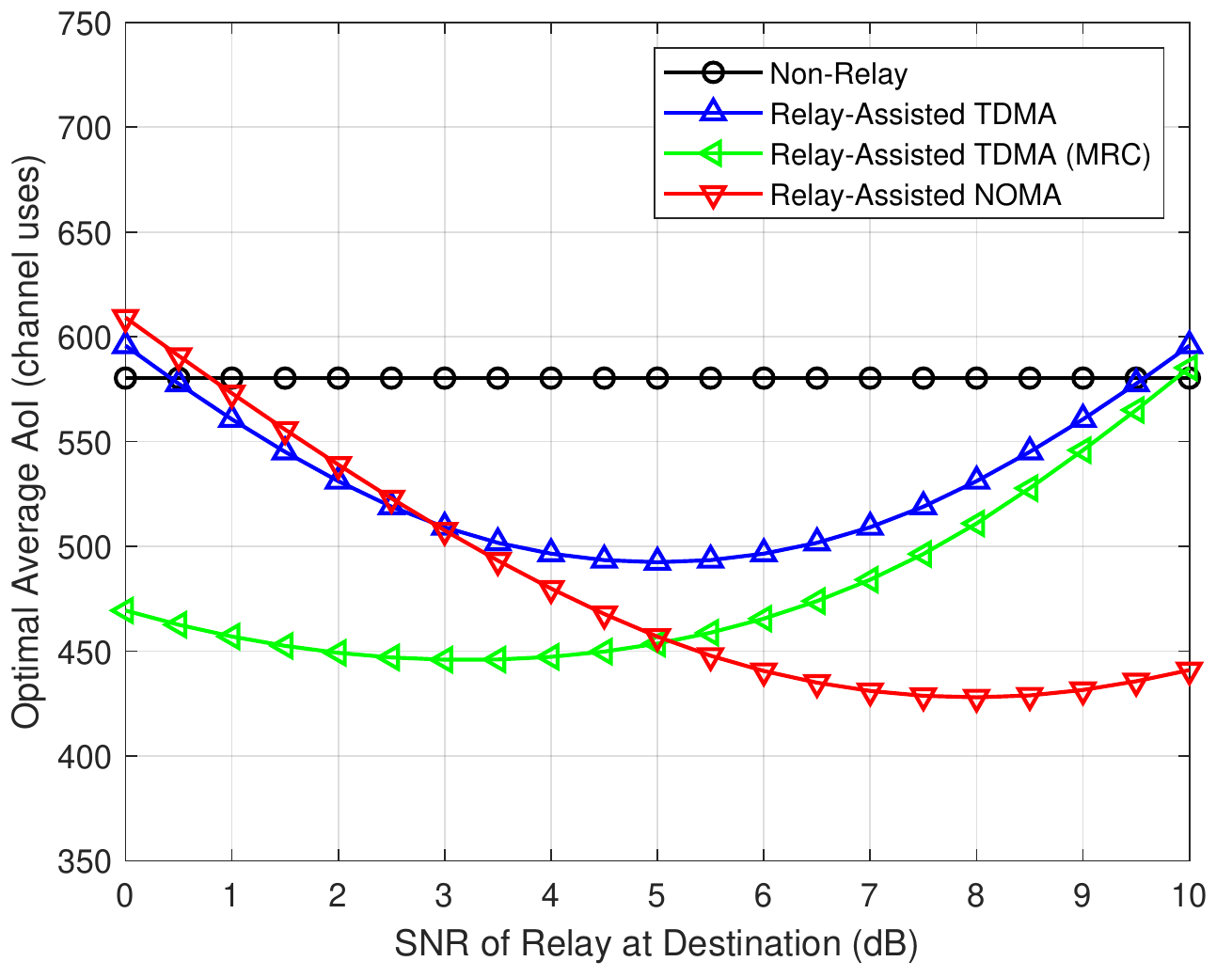}}
\caption{Performance comparison for the non-relay scheme and the relay-assisted TDMA (with and without MRC) and NOMA schemes: the average energy consumption versus ${\bar \gamma _{{\rm{rd}}}}$, when ${\bar \gamma _{{\rm{sd}}}}=-2dB$. The payload of an update packet is 128 bits.}
\label{fig-mrc}
\end{figure}

Fig. \ref{fig-mrc} plots the optimal average AoI performance of the non-relay scheme and the relay-assisted TDMA and NOMA schemes, versus ${\bar \gamma _{{\rm{rd}}}}$ with ${\bar \gamma _{{\rm{sr}}}} + {\bar \gamma _{{\rm{rd}}}} = 10dB$, when ${\bar \gamma _{{\rm{sd}}}}$ is $-2dB$. The payload of an update packet is $128$ bits. As expected, using MRC reduces the average AoI little when ${\bar \gamma _{{\rm{sr}}}}$ (${p_{{\rm{sr}}}}$) is small. This is because the relay can hardly decode the update packet sent by the source, and most of the time MRC cannot be used. When ${\bar \gamma _{{\rm{sr}}}}$ is large (${\bar \gamma _{{\rm{rd}}}}$ is small) and MRC is used, the average AoI performance of the relay-assisted TDMA scheme outperforms the non-MRC counterpart. 

Comparing the average AoI between TDMA and NOMA schemes, we see in Fig. \ref{fig-mrc} that when ${\bar \gamma _{{\rm{rd}}}}$ is small, the TDMA scheme with MRC gives a lower average AoI than the NOMA scheme. When ${\bar \gamma _{{\rm{rd}}}}$ increases, the relay-assisted NOMA scheme gradually outperforms its TDMA counterpart. The phenomenon and reasons herein are the same as the non-MRC scenario presented in Section \ref{section-aoI-comparison}. Overall, using MRC improves the average AoI performance when the SNR of the relay at the destination is low. 

\begin{figure*}[!t]
\setcounter{mytempeqncnt}{\value{equation}}
\footnotesize
\begin{align}
\mathbb{E}\left[ Z \right] &= \frac{{{\pi _{{S^ + }}}}}{{{\pi _{{S^ + }}} + {\pi _{{J^ + }}} + {\pi _{{S^R}}}}}{M_{{S^ + }V}} + \frac{{{\pi _{{J^ + }}}}}{{{\pi _{{S^ + }}} + {\pi _{{J^ + }}} + {\pi _{{S^R}}}}}{M_{{J^ + }V}} + \frac{{{\pi _{{S^R}}}}}{{{\pi _{{S^ + }}} + {\pi _{{J^ + }}} + {\pi _{{S^R}}}}}{M_{{S^R}V}}.
\label{f-E_Z-NOMA-equation}\\
\mathbb{E}\left[ {{Z^2}} \right] &= \frac{{{\pi _{{S^ + }}}}}{{{\pi _{{S^ + }}} + {\pi _{{J^ + }}} + {\pi _{{S^R}}}}}{N_{{S^ + }V}} + \frac{{{\pi _{{J^ + }}}}}{{{\pi _{{S^ + }}} + {\pi _{{J^ + }}} + {\pi _{{S^R}}}}}{N_{{J^ + }V}} + \frac{{{\pi _{{S^R}}}}}{{{\pi _{{S^ + }}} + {\pi _{{J^ + }}} + {\pi _{{S^R}}}}}{N_{{S^R}V}}.
\label{f-E_Z^2-NOMA-equation} \\
\mathbb{E}\left[ {\tau Z} \right] &= \left( {\frac{{{\pi _{{S^ + }}}}}{{{\pi _{{S^ + }}} + {\pi _{{J^ + }}} + {\pi _{{S^R}}}}}{M_{{S^ + }V}} + \frac{{{\pi _{{J^ + }}}}}{{{\pi _{{S^ + }}} + {\pi _{{J^ + }}} + {\pi _{{S^R}}}}}{M_{{J^ + }V}}} \right) + 2\left( {\frac{{{\pi _{{S^R}}}}}{{{\pi _{{S^ + }}} + {\pi _{{J^ + }}} + {\pi _{{S^R}}}}}{M_{{S^R}V}}} \right).
\label{f-E_tZ-NOMA-equation}
\end{align}
\hrulefill
\end{figure*}

\section{Proof of Proposition 2: The Average AoI of the Relay-assisted NOMA Scheme} \label{sec:appendix_B}
As in the TDMA scheme, to compute the average AoI of the relay-assisted NOMA scheme, we need to compute $\mathbb{E}\left[ Z \right]$, $\mathbb{E}\left[ {{Z^2}} \right]$, and $\mathbb{E}\left[ {\tau Z} \right]$ for the NOMA scheme. Following the same manner in Section \ref{section-OMA}, they are computed by (\ref{f-E_Z-NOMA-equation}), (\ref{f-E_Z^2-NOMA-equation}), and (\ref{f-E_tZ-NOMA-equation}), where the embedded components are computed in the following.

\noindent \textbf{Computation of ${\pi _{{S^ + }}},{\pi _{{J^ + }}},{\pi _{{S^R}}}$:} The stationary distribution $\pi$ of the Markov chain depicted in Fig.~\ref{fig-NOMA-chain} can be obtained by solving the equation $\pi {\Omega _{_{{\rm{NOMA}}}}} = \pi$, i.e.,
\begin{align}
    \pi  &= \left( {{\pi _{{S^ + }}},{\pi _{{J^ + }}},{\pi _{{S^R}}},{\pi _{{J^ - }}},{\pi _{{S^ - }}}} \right) \notag \\ &= (\frac{{\left( {1 - {p_{{\rm{sr}}}}} \right){p_{{\rm{sd}}}} + {p_{{\rm{sr}}}}{{\left( {{p_{{\rm{sd}}}}} \right)}^2} + {p_{{\rm{sr}}}}p_{{\rm{sd}}}^{\rm{N}}\left( {1 - {p_{{\rm{sd}}}}} \right)}}{{1 + {p_{{\rm{sr}}}}}}, \frac{{{p_{{\rm{sr}}}}{p_{{\rm{sd}}}}}}{{1 + {p_{{\rm{sr}}}}}}, \notag \\ &~~~\frac{{{p_{{\rm{sr}}}}\left( {1 - {p_{{\rm{sd}}}}} \right)\left( {p_{{\rm{rd}}}^{\rm{N}} - p_{{\rm{sd}}}^{\rm{N}}} \right)}}{{1 + {p_{{\rm{sr}}}}}}, \frac{{{p_{{\rm{sr}}}}\left( {1 - {p_{{\rm{sd}}}}} \right)}}{{1 + {p_{{\rm{sr}}}}}}, \notag \\ &~~~\frac{{\left( {1 - {p_{{\rm{sd}}}}} \right)\left( {1 + {p_{{\rm{sr}}}}{p_{{\rm{sd}}}} - {p_{{\rm{sr}}}}p_{{\rm{rd}}}^{\rm{N}}} \right)}}{{1 + {p_{{\rm{sr}}}}}}).
    \label{f-pi-NOMA}
\end{align}

\noindent \textbf{Computation of ${M_{aV}}$ and ${N_{aV}}$:} Following (\ref{f-M}), we can also compute ${M_{aV}}$ for the NOMA scheme. Based on the Markov chain in Fig.~\ref{fig-NOMA-chain}, the linear equations of ${M_{aV}}$ with different $a$ are
\begin{align}
\left\{ \begin{array}{l}
{M_{{S^ + }V}} = {\omega _{{S^ + }{S^ + }}} + {\omega _{{S^ + }{J^ + }}} + {\omega _{{S^ + }{J^ - }}}\left( {1 + {M_{{J^ - }V}}} \right) \\
~~~~~~~~~~~+ {\omega _{{S^ + }{S^ - }}}\left( {1 + {M_{{S^ - }V}}} \right)\\
{M_{{J^ + }V}} = {\omega _{{J^ + }{S^ + }}} + {\omega _{{J^ + }{S^ - }}}\left( {1 + {M_{{S^ - }V}}} \right)\\
{M_{{S^R}V}} = {\omega _{{S^R}{S^ + }}} + {\omega _{{S^R}{J^ + }}} + {\omega _{{S^R}{J^ - }}}\left( {1 + {M_{{J^ - }V}}} \right) \\
~~~~~~~~~~~+ {\omega _{{S^R}{S^ - }}}\left( {1 + {M_{{S^ - }V}}} \right)\\
{M_{{J^ - }V}} = {\omega _{{J^ - }{S^ + }}} + {\omega _{{J^ - }{S^R}}} + {\omega _{{J^ - }{S^ - }}}\left( {1 + {M_{{S^ - }V}}} \right)\\
{M_{{S^ - }V}} = {\omega _{{S^ - }{S^ + }}} + {\omega _{{S^ - }{J^ + }}} + {\omega _{{S^ - }{J^ - }}}\left( {1 + {M_{{J^ - }V}}} \right) \\
~~~~~~~~~~~+ {\omega _{{S^ - }{S^ - }}}\left( {1 + {M_{{S^ - }V}}} \right)
\end{array} \right..
\label{f-M-NOMA-equation}
\end{align}
Solving (\ref{f-M-NOMA-equation}), ${M_{aV}}$ where $a = \{ {S^ + },{J^ + },{S^R}\} $ is computed by
\begin{align}
\left\{ \begin{array}{l}
{M_{{S^ + }V}} = {M_{{S^R}V}} = \frac{{1 + {p_{{\rm{sr}}}}\left( {1 - {p_{{\rm{sd}}}}} \right)}}{{{p_{{\rm{sd}}}} + {p_{{\rm{sr}}}}p_{{\rm{rd}}}^{\rm{N}}\left( {1 - {p_{{\rm{sd}}}}} \right)}}\\
{M_{{J^ + }V}} = 1 + \left( {1 - {p_{{\rm{sd}}}}} \right){M_{{S^ + }V}}
\end{array} \right. .
\label{f-M-NOMA}
\end{align}

Similarly, we compute ${N_{aV}}$ based on equation (\ref{f-N}). Specifically, ${N_{aV}}, \ a = \{ {S^ + },{J^ + },{S^R}\}, $ is obtained by solving the linear equations of ${N_{aV}}$ with different $a$,
\begin{align}
\left\{ \begin{array}{l}
{N_{{S^ + }V}} = {\omega _{{S^ + }{S^ + }}} + {\omega _{{S^ + }{J^ + }}} + {\omega _{{S^ + }{J^ - }}}\left( {1 + {N_{{J^ - }V}} + 2{M_{{J^ - }V}}} \right) \\
~~~~~~~~~~~+ {\omega _{{S^ + }{S^ - }}}\left( {1 + {N_{{S^ - }V}} + 2{M_{{S^ - }V}}} \right)\\
{N_{{J^ + }V}} = {\omega _{{J^ + }{S^ + }}} + {\omega _{{J^ + }{S^ - }}}\left( {1 + {N_{{S^ - }V}} + 2{M_{{S^ - }V}}} \right)\\
{N_{{S^R}V}} = {\omega _{{S^R}{S^ + }}} + {\omega _{{S^R}{J^ + }}} + {\omega _{{S^R}{J^ - }}}\left( {1 + {N_{{J^ - }V}} + 2{M_{{J^ - }V}}} \right) \\
~~~~~~~~~~~+ {\omega _{{S^R}{S^ - }}}\left( {1 + {N_{{S^ - }V}} + 2{M_{{S^ - }V}}} \right)\\
{N_{{J^ - }V}} = {\omega _{{J^ - }{S^ + }}} + {\omega _{{J^ - }{S^R}}} + {\omega _{{J^ - }{S^ - }}}\left( {1 + {N_{{S^ - }V}} + 2{M_{{S^ - }V}}} \right)\\
{N_{{S^ - }V}} = {\omega _{{S^ - }{S^ + }}} + {\omega _{{S^ - }{J^ + }}} + {\omega _{{S^ - }{J^ - }}}\left( {1 + {N_{{J^ - }V}} + 2{M_{{J^ - }V}}} \right) \\
~~~~~~~~~~~+ {\omega _{{S^ - }{S^ - }}}\left( {1 + {N_{{S^ - }V}} + 2{M_{{S^ - }V}}} \right)
\end{array} \right..
\label{f-N-NOMA-equation}
\end{align}
Hence, we have
\begin{align}
\left\{ \begin{array}{l}
{N_{{S^ + }V}} = {N_{{S^R}V}} = \frac{{1 + 3{p_{{\rm{sr}}}}\left( {1 - {p_{{\rm{sd}}}}} \right)}}{{{p_{{\rm{sd}}}} + {p_{{\rm{sr}}}}p_{{\rm{rd}}}^{\rm{N}}\left( {1 - {p_{{\rm{sd}}}}} \right)}}\\
~~~~~~~~~~~~~~~~~~~
+ \frac{{2\left[ {1 + {p_{{\rm{sr}}}}\left( {1 - {p_{{\rm{sd}}}}} \right)} \right]\left[ {\left( {1 - {p_{{\rm{sd}}}}} \right)\left( {1 + {p_{{\rm{sr}}}} - 2{p_{{\rm{sr}}}}p_{{\rm{rd}}}^{\rm{N}}} \right)} \right]}}{{{{\left[ {{p_{{\rm{sd}}}} + {p_{{\rm{sr}}}}p_{{\rm{rd}}}^{\rm{N}}\left( {1 - {p_{{\rm{sd}}}}} \right)} \right]}^2}}}\\
{N_{{J^ + }V}} = 1 + \left( {1 - {p_{{\rm{sd}}}}} \right)\left[ {{N_{{S^ + }V}} + 2{M_{{S^ + }V}}} \right]
\end{array} \right. .
\label{f-N-NOMA}
\end{align}

\begin{figure*}[!t]
\setcounter{mytempeqncnt}{\value{equation}}
\footnotesize
\begin{align}
\mathbb{E}\left[ Z \right] &= \frac{{1 + {p_{{\rm{sr}}}}}}{{{p_{{\rm{sd}}}} + {p_{{\rm{sr}}}}p_{{\rm{rd}}}^{\rm{N}} + {p_{{\rm{sr}}}}{p_{{\rm{sd}}}}\left( {{p_{{\rm{sd}}}} - p_{{\rm{rd}}}^{\rm{N}}} \right)}}.
\label{f-E_Z-NOMA} \\
\mathbb{E}\left[ {{Z^2}} \right] &= \frac{{1 + {p_{{\rm{sr}}}}\left( {3 - 2{p_{{\rm{sd}}}}} \right)}}{{{p_{{\rm{sd}}}} + {p_{{\rm{sr}}}}p_{{\rm{rd}}}^{\rm{N}} + {p_{{\rm{sr}}}}{p_{{\rm{sd}}}}\left( {{p_{{\rm{sd}}}} - p_{{\rm{rd}}}^{\rm{N}}} \right)}} 
+ \frac{{2\left[ {1 + {p_{{\rm{sr}}}}\left( {1 - {p_{{\rm{sd}}}}} \right)} \right]\left[ {\left( {1 - {p_{{\rm{sd}}}}} \right)\left( {1 + {p_{{\rm{sr}}}} - 2{p_{{\rm{sr}}}}p_{{\rm{rd}}}^{\rm{N}}} \right) + {p_{{\rm{sr}}}}{p_{{\rm{sd}}}}\left( {1 - {p_{{\rm{sd}}}}} \right)} \right]}}{{\left[ {{p_{{\rm{sd}}}} + {p_{{\rm{sr}}}}p_{{\rm{rd}}}^{\rm{N}} + {p_{{\rm{sr}}}}{p_{{\rm{sd}}}}\left( {{p_{{\rm{sd}}}} - p_{{\rm{rd}}}^{\rm{N}}} \right)} \right]\left[ {{p_{{\rm{sd}}}} + {p_{{\rm{sr}}}}p_{{\rm{rd}}}^{\rm{N}}\left( {1 - {p_{{\rm{sd}}}}} \right)} \right]}}.
\label{f-E_Z^2-NOMA} \\
\mathbb{E}\left[ {\tau Z} \right] &= \frac{{1 + {p_{{\rm{sr}}}}}}{{{p_{{\rm{sd}}}} + {p_{{\rm{sr}}}}p_{{\rm{rd}}}^{\rm{N}} + {p_{{\rm{sr}}}}{p_{{\rm{sd}}}}\left( {{p_{{\rm{sd}}}} - p_{{\rm{rd}}}^{\rm{N}}} \right)}}
 + \frac{{{p_{{\rm{sr}}}}\left( {1 - {p_{{\rm{sd}}}}} \right)\left( {p_{{\rm{rd}}}^{\rm{N}} - p_{{\rm{sd}}}^{\rm{N}}} \right)\left[ {1 + {p_{{\rm{sr}}}}\left( {1 - {p_{{\rm{sd}}}}} \right)} \right]}}{{\left[ {{p_{{\rm{sd}}}} + {p_{{\rm{sr}}}}p_{{\rm{rd}}}^{\rm{N}} + {p_{{\rm{sr}}}}{p_{{\rm{sd}}}}\left( {{p_{{\rm{sd}}}} - p_{{\rm{rd}}}^{\rm{N}}} \right)} \right]\left[ {{p_{{\rm{sd}}}} + {p_{{\rm{sr}}}}p_{{\rm{rd}}}^{\rm{N}}\left( {1 - {p_{{\rm{sd}}}}} \right)} \right]}}.
\label{f-E_tZ-NOMA} 
\end{align}
\hrulefill
\end{figure*}

Therefore, using the components computed above, $\mathbb{E}\left[ Z \right]$, $\mathbb{E}\left[ {{Z^2}} \right]$, and $\mathbb{E}\left[ {\tau Z} \right]$ are given by (\ref{f-E_Z-NOMA}), (\ref{f-E_Z^2-NOMA}), and (\ref{f-E_tZ-NOMA}). Finally, we can compute ${\bar \Delta _{{\rm{NOMA}}}}$ in (\ref{f-aoi-NOMA}) by substituting (\ref{f-E_Z-NOMA}), (\ref{f-E_Z^2-NOMA}), and (\ref{f-E_tZ-NOMA}) into  (\ref{f-average-AoI-s}).

\section{Computations of $p_{{\rm{rd}}}^{\rm{N}}$ and $p_{{\rm{sd}}}^{\rm{N}}$} \label{sec:appendix_C}

In the NOMA relay-assisted scheme, when SIC is used and the signal from the source to the destination is treated as noise, based on \cite{Le2022}, the CDF of the effective SNR $\gamma _{{\rm{rd}}}^{\rm{N}}$ from the relay to the destination is given by 
\begin{align}
{F_{\gamma _{{\rm{rd}}}^{\rm{N}}}}\left( x \right) &= 1 - \left( {\int_0^\infty  {{e^{ - t}}dt} } \right){e^{ - \frac{x}{{{{\bar \gamma }_{{\rm{rd}}}}}}}} \times \frac{{\int_0^\infty  {{e^{ - t}}dt} }}{{1 + \frac{{{{\bar \gamma }_{{\rm{sd}}}}}}{{{{\bar \gamma }_{{\rm{rd}}}}}}x}} \notag \\
&= 1 - \frac{{{e^{ - \frac{x}{{{{\bar \gamma }_{{\rm{rd}}}}}}}}}}{{1 + \frac{{{{\bar \gamma }_{{\rm{sd}}}}}}{{{{\bar \gamma }_{{\rm{rd}}}}}}x}}
\end{align}
Therefore, $p_{{\rm{rd}}}^{\rm{N}}$ (\ref{f-p_rd_noma}) can be computed by $p_{{\rm{rd}}}^{\rm{N}} = 1 - \bar \varepsilon _{{\rm{rd}}}^{\rm{N}}$, where $\bar \varepsilon _{{\rm{rd}}}^{\rm{N}}$ is computed by (\ref{f-per_rd_noma}). For $p_{{\rm{sd}}}^{\rm{N}}$ (\ref{f-p_sd_noma}), it can be computed by $p_{{\rm{sd}}}^{\rm{N}} = p_{{\rm{rd}}}^{\rm{N}}{p_{{\rm{sd}}}}$, i.e., perfect SIC is assumed so that the destination tries to decode an interference-free packet from the source.

\begin{figure*}[!t]
\setcounter{mytempeqncnt}{\value{equation}}
\footnotesize
\begin{align}
\bar \varepsilon _{{\rm{rd}}}^{\rm{N}} &= \int_0^\infty  {{{\left( {1 - \frac{{{e^{ - \frac{x}{{{{\bar \gamma }_{{\rm{rd}}}}}}}}}}{{1 + \frac{{{{\bar \gamma }_{{\rm{sd}}}}}}{{{{\bar \gamma }_{{\rm{rd}}}}}}x}}} \right)}^{'}}Q\left( {\frac{{\sqrt n \left( {\log \left( {1 + x} \right) - \frac{D}{n}} \right)}}{{\sqrt {1 - \frac{1}{{{{\left( {1 + x} \right)}^2}}}} }}} \right)} dx
= \int_0^{\delta  + \frac{1}{{2\beta }}} {{{\left( {1 - \frac{{{e^{ - \frac{x}{{{{\bar \gamma }_{{\rm{rd}}}}}}}}}}{{1 + \frac{{{{\bar \gamma }_{{\rm{sd}}}}}}{{{{\bar \gamma }_{{\rm{rd}}}}}}x}}} \right)}^{'}}} dx + \int_{\delta  + \frac{1}{{2\beta }}}^{\delta  - \frac{1}{{2\beta }}} {{{\left( {1 - \frac{{{e^{ - \frac{x}{{{{\bar \gamma }_{{\rm{rd}}}}}}}}}}{{1 + \frac{{{{\bar \gamma }_{{\rm{sd}}}}}}{{{{\bar \gamma }_{{\rm{rd}}}}}}x}}} \right)}^{'}}} \left[ {\beta \left( {x - \delta } \right) + \frac{1}{2}} \right]dx \notag \\
&= 1 - \frac{{{e^{ - \frac{{\delta  + \frac{1}{{2\beta }}}}{{{{\bar \gamma }_{{\rm{rd}}}}}}}}}}{{1 + \frac{{{{\bar \gamma }_{{\rm{sd}}}}}}{{{{\bar \gamma }_{{\rm{rd}}}}}}\left( {\delta  + \frac{1}{{2\beta }}} \right)}} + \left( {\frac{1}{2} - \beta \delta } \right)\left( {\frac{{{e^{ - \frac{{\delta  + \frac{1}{{2\beta }}}}{{{{\bar \gamma }_{{\rm{rd}}}}}}}}}}{{1 + \frac{{{{\bar \gamma }_{{\rm{sd}}}}}}{{{{\bar \gamma }_{{\rm{rd}}}}}}\left( {\delta  + \frac{1}{{2\beta }}} \right)}} - \frac{{{e^{ - \frac{{\delta  - \frac{1}{{2\beta }}}}{{{{\bar \gamma }_{{\rm{rd}}}}}}}}}}{{1 + \frac{{{{\bar \gamma }_{{\rm{sd}}}}}}{{{{\bar \gamma }_{{\rm{rd}}}}}}\left( {\delta  - \frac{1}{{2\beta }}} \right)}}} \right) + \beta \int_{\delta  + \frac{1}{{2\beta }}}^{\delta  - \frac{1}{{2\beta }}} {xd\left( {1 - \frac{{{e^{ - \frac{x}{{{{\bar \gamma }_{{\rm{rd}}}}}}}}}}{{1 + \frac{{{{\bar \gamma }_{{\rm{sd}}}}}}{{{{\bar \gamma }_{{\rm{rd}}}}}}x}}} \right)} \notag \\
&= 1 + \beta \int_{\delta  + \frac{1}{{2\beta }}}^{\delta  - \frac{1}{{2\beta }}} {\frac{{{e^{ - \frac{x}{{{{\bar \gamma }_{{\rm{rd}}}}}}}}}}{{1 + \frac{{{{\bar \gamma }_{{\rm{sd}}}}}}{{{{\bar \gamma }_{{\rm{rd}}}}}}x}}dx} 
= 1 + \beta \frac{{{{\bar \gamma }_{{\rm{rd}}}}}}{{{{\bar \gamma }_{{\rm{sd}}}}}}\int_{\delta  + \frac{1}{{2\beta }}}^{\delta  - \frac{1}{{2\beta }}} {\frac{{{e^{ - \frac{x}{{{{\bar \gamma }_{{\rm{rd}}}}}}}}}}{{1 + \frac{{{{\bar \gamma }_{{\rm{sd}}}}}}{{{{\bar \gamma }_{{\rm{rd}}}}}}x}}d\left( {1 + \frac{{{{\bar \gamma }_{{\rm{sd}}}}}}{{{{\bar \gamma }_{{\rm{rd}}}}}}x} \right)} \notag \\
&= 1 + \beta \frac{{{{\bar \gamma }_{{\rm{rd}}}}}}{{{{\bar \gamma }_{{\rm{sd}}}}}}\int_{1 + \frac{{{{\bar \gamma }_{{\rm{sd}}}}}}{{{{\bar \gamma }_{{\rm{rd}}}}}}\left( {\delta  + \frac{1}{{2\beta }}} \right)}^{1 + \frac{{{{\bar \gamma }_{{\rm{sd}}}}}}{{{{\bar \gamma }_{{\rm{rd}}}}}}\left( {\delta  - \frac{1}{{2\beta }}} \right)} {\frac{{{e^{ - \frac{{t - 1}}{{{{\bar \gamma }_{{\rm{sd}}}}}}}}}}{t}dt} {\rm{  }}
= 1 + \beta \frac{{{{\bar \gamma }_{{\rm{rd}}}}}}{{{{\bar \gamma }_{{\rm{sd}}}}}}{e^{\frac{1}{{{{\bar \gamma }_{{\rm{sd}}}}}}}}\int_{\frac{1}{{{{\bar \gamma }_{{\rm{sd}}}}}} + \frac{{\delta  + \frac{1}{{2\beta }}}}{{{{\bar \gamma }_{{\rm{rd}}}}}}}^{\frac{1}{{{{\bar \gamma }_{{\rm{sd}}}}}} + \frac{{\delta  - \frac{1}{{2\beta }}}}{{{{\bar \gamma }_{{\rm{rd}}}}}}} {\frac{{{e^{ - x}}}}{x}dx} {\rm{ }}
= 1 + \beta \frac{{{{\bar \gamma }_{{\rm{rd}}}}}}{{{{\bar \gamma }_{{\rm{sd}}}}}}{e^{\frac{1}{{{{\bar \gamma }_{{\rm{sd}}}}}}}}\int_{ - \left( {\frac{1}{{{{\bar \gamma }_{{\rm{sd}}}}}} + \frac{{\delta  + \frac{1}{{2\beta }}}}{{{{\bar \gamma }_{{\rm{rd}}}}}}} \right)}^{ - \left( {\frac{1}{{{{\bar \gamma }_{{\rm{sd}}}}}} + \frac{{\delta  - \frac{1}{{2\beta }}}}{{{{\bar \gamma }_{{\rm{rd}}}}}}} \right)} {\frac{{{e^t}}}{t}dt} \notag \\
&= 1 + \beta \frac{{{{\bar \gamma }_{{\rm{rd}}}}}}{{{{\bar \gamma }_{{\rm{sd}}}}}}{e^{\frac{1}{{{{\bar \gamma }_{{\rm{sd}}}}}}}}\left( {Ei\left( { - \left( {\frac{1}{{{{\bar \gamma }_{{\rm{sd}}}}}} + \frac{{\delta  - \frac{1}{{2\beta }}}}{{{{\bar \gamma }_{{\rm{rd}}}}}}} \right)} \right) - Ei\left( { - \left( {\frac{1}{{{{\bar \gamma }_{{\rm{sd}}}}}} + \frac{{\delta  + \frac{1}{{2\beta }}}}{{{{\bar \gamma }_{{\rm{rd}}}}}}} \right)} \right)} \right)
\label{f-per_rd_noma}
\\p_{{\rm{sd}}}^{\rm{N}} &= p_{{\rm{rd}}}^{\rm{N}}{p_{{\rm{sd}}}} = {\beta ^2}{\bar \gamma _{{\rm{rd}}}}{e^{\frac{1}{{{{\bar \gamma }_{{\rm{sd}}}}}}}}\left( {{e^{ - \frac{1}{{{{\bar \gamma }_{{\rm{sd}}}}}}\left( {\delta  - \frac{1}{{2\beta }}} \right)}} - {e^{ - \frac{1}{{{{\bar \gamma }_{{\rm{sd}}}}}}\left( {\delta  + \frac{1}{{2\beta }}} \right)}}} \right)\left( {Ei\left( { - \left( {\frac{1}{{{{\bar \gamma }_{{\rm{sd}}}}}} + \frac{{\delta  + \frac{1}{{2\beta }}}}{{{{\bar \gamma }_{{\rm{rd}}}}}}} \right)} \right) - Ei\left( { - \left( {\frac{1}{{{{\bar \gamma }_{{\rm{sd}}}}}} + \frac{{\delta  - \frac{1}{{2\beta }}}}{{{{\bar \gamma }_{{\rm{rd}}}}}}} \right)} \right)} \right)
\end{align}
\hrulefill
\end{figure*}

\end{document}